\documentclass[aps,prx,twocolumn,showpacs,superscriptaddress,longbibliography]{revtex4-2}
\usepackage{amsmath,amssymb,graphicx}
\usepackage{graphicx,epstopdf}
\usepackage{gensymb}
\usepackage[dvipsnames]{xcolor}
\usepackage[T1]{fontenc}
\usepackage[utf8]{inputenc}
\usepackage{lipsum}
\usepackage{adjustbox}
\newlength\mylength
\newcommand{\be}{\begin{equation}}
\newcommand{\ee}{\end{equation}}
\newcommand{\bea}{\begin{eqnarray}}
\newcommand{\eea}{\end{eqnarray}}
\newcommand{\bse}{\begin{subequations}}
\newcommand{\ese}{\end{subequations}}

\usepackage{multibib}
\usepackage{color}
\usepackage[colorlinks,bookmarks=false,citecolor=darkblue,linkcolor=red,urlcolor=blue]{hyperref}

\definecolor{darkred}{rgb}{0.7,0.0,0.0}

\definecolor{darkblue}{rgb}{0,0.02,0.45}

\definecolor{darkgreen}{rgb}{0.02,0.45,0.0}

\definecolor{violet}{rgb}{0.8,0.2,0.6}

\begin{document}

\title{Double magnetic transition, complex field-induced phases, and large magnetocaloric effect in the frustrated garnet compound Mn$_{3}$Cr$_{2}$Ge$_{3}$O$_{12}$}
\author{S. Mohanty}
\thanks{These authors have equal contribution.}
\author{A. Magar}
\thanks{These authors have equal contribution.}
\author{Vikram Singh}
\author{S. S. Islam}
\author{S.~Guchhait}
\affiliation{School of Physics, Indian Institute of Science Education and Research Thiruvananthapuram-695551, India}
\author{A. Jain}
\author{S. M. Yusuf}
\affiliation{Solid State Physics Division, Bhabha Atomic Research Centre, Mumbai 400 085, India}
\author{A. A. Tsirlin}
\affiliation{Felix Bloch Institute for Solid-State Physics, Leipzig University, 04103 Leipzig, Germany}
\author{R. Nath}
\email{rnath@iisertvm.ac.in}
\affiliation{School of Physics, Indian Institute of Science Education and Research Thiruvananthapuram-695551, India}
\date{\today}

\begin{abstract}
A detailed study of the magnetic and magnetocaloric properties of a garnet compound Mn$_{3}$Cr$_{2}$Ge$_{3}$O$_{12}$ is carried out using x-ray diffraction, magnetization, heat capacity, and neutron diffraction measurements as well as \textit{ab initio} band-structure calculations. This compound manifests two successive magnetic transitions at $T_{\rm N1} \simeq 4.5$~K and $T_{\rm N2} \simeq 2.7$~K. Neutron powder diffraction experiments reveal that these two transitions correspond to the collinear and non-collinear antiferromagnetic ordering of the nonfrustrated Cr$^{3+}$ and frustrated Mn$^{2+}$ sublattices, respectively. The interactions within each of the Cr and Mn sublattices are antiferromagnetic, while the inter-sublattice interactions are ferromagnetic. The $H-T$ phase diagram is quite complex and displays multiple phases under magnetic field, which can be attributed to the frustrated nature of the spin lattice. Mn$_{3}$Cr$_{2}$Ge$_{3}$O$_{12}$ shows a large magnetocaloric effect with a maximum value of isothermal entropy change $\Delta S_{\rm m} \simeq -23$~J/kg-K and adiabatic temperature change $\Delta T_{\rm ad} \simeq 9$~K for a field change of 7~T. Further, a large value of the relative cooling power ($RCP \simeq 360$~J/kg) demonstrates the promise of using this compound in magnetic refrigeration.
\end{abstract}

\maketitle

\section{Introduction}
Frustrated magnets are widely studied because of their potential to host a variety of exotic ground states~\cite{Starykh052502,Ramirez453}. In particular, the geometrically frustrated magnets in three dimensions (3D) that include pyrochlore and hyperkagome lattices made up of corner-sharing tetrahedra and corner-sharing triangles, respectively, have very intricate ground states~\cite{Gardner53,Jin054408}.
Garnet is a family of compounds with the general formula $A_3B_2C_3\rm{O}_{12}$
which can accommodate a large variety of chemical constituents. Here, $A$, $B$, and $C$ occupy the dodecahedral, octahedral, and tetrahedral crystallographic sites, respectively~\cite{Geller1}. The garnet family provides a convenient platform to observe wide variety of non-trivial properties by introducing different (3$d$ and 4$f$) magnetic ions at different crystallographic sites. The magnetic ions present only at the $A$-sites form a geometrically frustrated hyperkagome lattice, giving rise to complex magnetic structures in Co$_3$Al$_2$Si$_3$O$_{12}$~\cite{Cui144424} or magnetoelectric effect in Mn$_3$Al$_2$Ge$_3$O$_{12}$~\cite{Min86}. 
Similarly, $4f$ ions occupying the $A$-sites show very peculiar low-temperature features. For example, the celebrated garnet compound Gd$_3$Ga$_5$O$_{12}$ manifests a spin liquid with a hidden long-range order (LRO)~\cite{Paddison179}, Tb$_3$Ga$_5$O$_{12}$ exhibits a field-induced LRO~\cite{Kamazawa064412,Wawrzy094442}, Ho$_3$Ga$_5$O$_{12}$ shows a disordered ground state, etc~\cite{Zhou140406}.

Another class of garnets can be obtained by introducing either same or different magnetic ions at both $B$ and $C$ sites. One such family is $R_3$Fe$_5$O$_{12}$ (where $R$ is a rare-earth ion) that received wide attention because of the ferrimagnetic ground state~\cite{Konstantin149} as in Y$_3$Fe$_5$O$_{12}$ that additionally displays magnetoelectric effect and thermal spin dynamics~\cite{Kohara104419,Barker217201}. In these garnet compounds, each of the $B$ and $C$ sublattices is ferromagnetic (FM), whereas antiferromagnetic (AFM) interactions between the sublattices give rise to the ferrimagnetic order. An opposite situation can be envisaged in garnets with dissimilar magnetic ions occupying two sublattices where a different coupling regime with the increased frustration could be realized, especially if intra-sublattice interactions are AFM in nature. In this category, only a few compounds are reported with very preliminary magnetic measurements. For instance, Mn$_3$Fe$_2$Ge$_3$O$_{12}$ undergoes an AFM ordering at around 6~K~\cite{Bozorth263}, whereas Mn$_{3}$Cr$_{2}$Ge$_{3}$O$_{12}$ shows two subsequent AFM orderings at low temperatures~\cite{Belov173,Valyanskaya2114}.

Owing to their large magnetocaloric effect (MCE), frustrated magnets are considered as promising materials for magnetic refrigeration~\cite{Tokiwa42} that uses adiabatic demagnetization technique to achieve low temperatures. It is an environment-friendly replacement for gas compression technique implemented in standard refrigerators for room-temperature applications and a cost-effective replacement to achieve sub-Kelvin temperatures over expensive $^3$He and $^4$He. To attain low temperatures using MCE, magnets with low transition temperatures and large entropy changes are desirable. Magnetic frustration can impede the magnetic ordering and enhance MCE, for example in the garnet family~\cite{Zhitomirsky104421}. Indeed, Gd$_3$Ga$_5$O$_{12}$ or gadolinium gallium garnet (GGG) exhibits a giant MCE with the isothermal entropy change of $\sim 450$~J/kg K and base temperatures as low as 800~mK~\cite{Kleinhans014038}. Therefore, GGG is commercially used in magnetic refrigerators. Similarly, few other compounds of the garnet family are reported to show large cooling power for use in the milli-Kelvin range~\cite{Kleinhans014038}. Garnets also proved to be very useful in technological applications. For instance, neodymium-doped garnets are good laser materials, ferrimagnetic garnets have applications in electronic devices etc~\cite{Urata801}.

\begin{figure*}
\includegraphics[width=\textwidth]{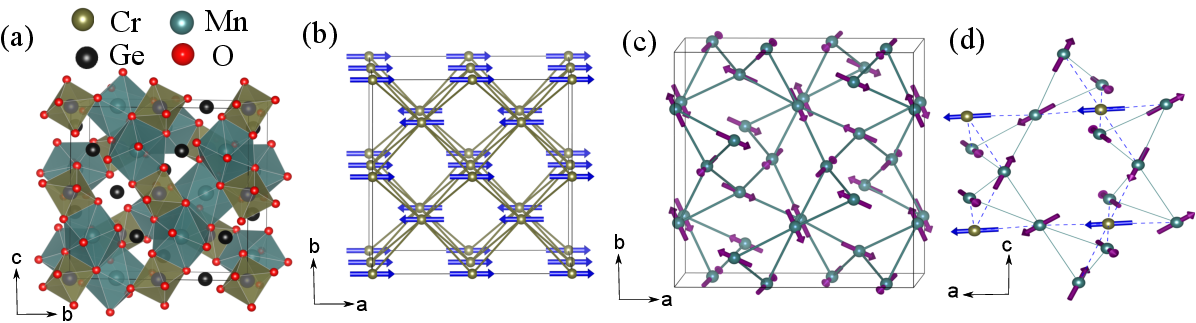}
\caption{\label{Fig1} (a) Three-dimensional view of the crystal structure of MCGO. (b) Arrangement of spins in the Cr$^{3+}$ sublattice. (c) Hyperkagome structure formed by the Mn$^{2+}$ ions. (d) Coupling between Mn$^{2+}$ and Cr$^{3+}$ ions that yields pyrochlore-like structure.}
\end{figure*}
In the present work, we re-visited the magnetic properties of the garnet compound Mn$_{3}$Cr$_{2}$Ge$_{3}$O$_{12}$ (MCGO) in detail by means of magnetic, thermodynamic, neutron diffraction, and magnetocaloric measurements.
MCGO is reported to crystallize in a cubic structure with the space group $Ia\bar{3}d$ (No.~230) at room temperature~\cite{Lippi35}. Here, Mn$^{2+}$ ion is situated in a dodecahedral site coordinated with eight oxygen atoms, Cr$^{3+}$ is forming octahedra with six oxygen atoms, and Ge$^{4+}$ is forming slightly distorted tetrahedra with four oxygen atoms as shown in Fig.~\ref{Fig1}(a). The CrO$_6$ octahedra are corner-shared with the GeO$_4$ tetrahedra making a 3D structure with the shortest Cr--Cr distance of $\sim 5.196$~\AA. Similarly, the MnO$_8$ units are directly edge-shared to make a frustrated hyper-kagome lattice with the shortest Mn--Mn distance of $\sim 3.674$~\AA, though additional interactions via GeO$_4$ may also be possible.
The Mn$^{2+}$ and Cr$^{3+}$ sublattices are further coupled with each other as shown in Fig.~\ref{Fig1}(d), resulting in a pyrochlore-like structure with the shortest Mn$^{2+}$--Cr$^{3+}$ distance of $\sim 3.354$~\AA. Our measurements reveal double magnetic transition in zero field and a complex low-temperature phase diagram in the applied field. Zero-field magnetic structures are determined via powder neutron diffraction experiments. Moreover, a large MCE is obtained across the transitions.

\section{Methods}
\begin{figure}
\includegraphics[width=\columnwidth]{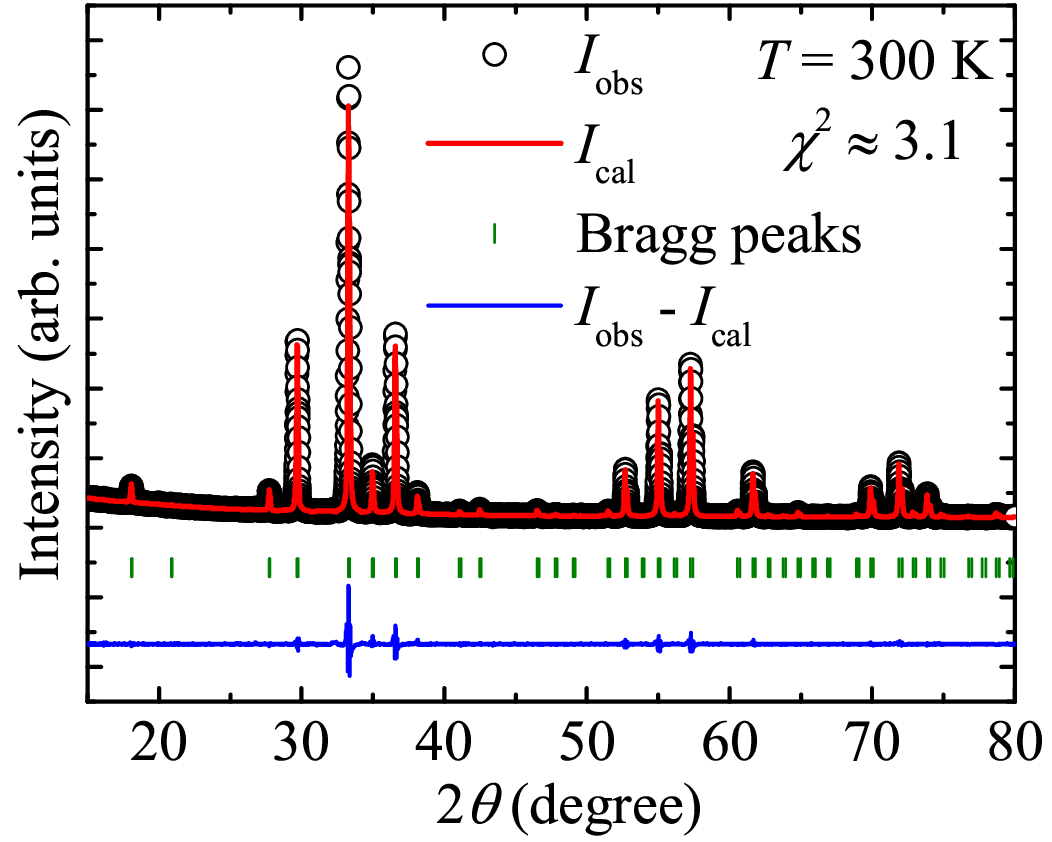}
\caption{\label{Fig2} Powder XRD data collected at room temperature. The red solid line is the Rietveld fit to the data, the vertical bars are the Bragg-peak positions, and the blue line at the bottom is the difference between the experimentally observed and calculated intensities.}
\end{figure}
Polycrystalline samples of MCGO were prepared using the conventional solid-state reaction technique by heating stoichiometric mixtures of Mn$_3$O$_4$ (Aldrich, 99.99\%), Cr$_2$O$_3$ (Aldrich, 99.99\%), and GeO$_2$ (Aldrich, 99.99\%). These reagents were finely ground, pressed into pellets, and fired at $800-1200 \degree$C with multiple intermediate regrindings. Finally, green-colored polycrystalline sample of MCGO was obtained. The phase purity of the product was confirmed by powder x-ray diffraction (XRD) recorded at room temperature using a PANalytical x-ray diffractometer (Cu\textit{K$_{\alpha}$} radiation, $\lambda_{\rm avg}\simeq 1.5418$~\AA). Figure~\ref{Fig2} presents the powder XRD pattern of MCGO along with the Rietveld fit. With the help of Rietveld refinement, all the diffraction peaks of MCGO could be indexed with the cubic unit cell [$Ia\bar{3}d$ (No.~230)], taking the initial structural parameters from Ref.~\cite{Lippi35}. The absence of any unidentified peak suggests the phase purity of the polycrystalline sample. The obtained lattice parameters at room temperature are $a = b = c \simeq 12.029(5)$~\AA~and unit-cell volume $V_{\rm cell} \simeq 1740.60(1)$~\AA$^3$, which are in close agreement with the previous report~\cite{Lippi35}.

Magnetization ($M$) measurements were performed as a function of temperature (0.4~K~$\leq T \leq$~380~K) and magnetic field (0~$\leq H \leq$~7~T) using a superconducting quantum interference device (SQUID) (MPMS-3, Quantum Design) magnetometer. Measurements below 1.8~K and down to 0.4~K were carried out using a $^{3}$He attachment to the MPMS. Heat capacity ($C_{\rm p}$) as a function of $T$ (0.4~K~$\leq T \leq$~250~K) and $H$ (0~$\leq H \leq 9$~T) was measured on a small piece of sintered pellet using the relaxation technique in the physical property measurement system (PPMS, Quantum Design). Measurements below 1.8~K were carried out using an additional $^{3}$He insert in the PPMS.

To solve the magnetic structure, temperature-dependent (2.5~K~$\leq T \leq 300$~K) neutron powder diffraction (NPD) experiments were carried out using the powder diffractometer at the Dhruva reactor, Bhaba Atomic Research Center (BARC), Mumbai, India. Measurements were carried out using the powder diffractometer PD-I ($\lambda\simeq 1.094$~\AA) with three linear position-sensitive detectors. 
The one-dimensional neutron-depolarization measurements were performed using the polarized neutron spectrometer (PNS) at the Dhruva reactor with a constant wavelength of $\lambda\simeq 1.205$~\AA. For these measurements, a Cu$_2$MnAl Heusler single crystal [(111) reflection] was used to produce the incident polarized neutron beams (along $z$-direction) and a Co$_{0.92}$Fe$_{0.8}$ [(200) reflection] single crystal was used to analyze the polarization of the transmitted (scattered) beam.  A $\pi$-flipper placed just before the sample allowed the polarization state of the neutron beam on the powder sample to be controlled between spin-up and spin-down states. The sample was placed in an Al sample holder. The flipping ratio of the beam was determined by measuring the intensities of neutrons in non-spin-flip and spin-flip channels with the $\pi$-flipper on and off, respectively. Rietveld refinement of the powder XRD and NPD data was performed using the \verb"FullProf" software package~\cite{Carvajal55}.

Magnetic couplings of MCGO were determined by density-functional (DFT) band-structure calculations using the mapping procedure~\cite{xiang2011}. The calculations were performed in the VASP code~\cite{vasp1,vasp2} with the Perdew-Burke-Ernzerhof exchange-correlation potential~\cite{pbe96} and up to 64 $k$-points in the first Brillouin zone. Correlation effects in the $3d$ shell were taken into account on the mean-field DFT+$U$ level with the on-site Coulomb repulsion $U_{\rm Mn}=5$\,eV~\cite{nath2014} and $U_{\rm Cr}=3$\,eV~\cite{janson2014}, as well as Hund's coupling $J_H=1$\,eV for both transition-metal atoms. The exchange coupling $J_i$ are calculated per bond and normalized to $S=\frac32$ for Cr$^{3+}$ and $S=\frac52$ for Mn$^{2+}$.

\section{Results and Discussion}

\subsection{Magnetization}
\begin{figure*}
\includegraphics[width=0.8\textwidth]{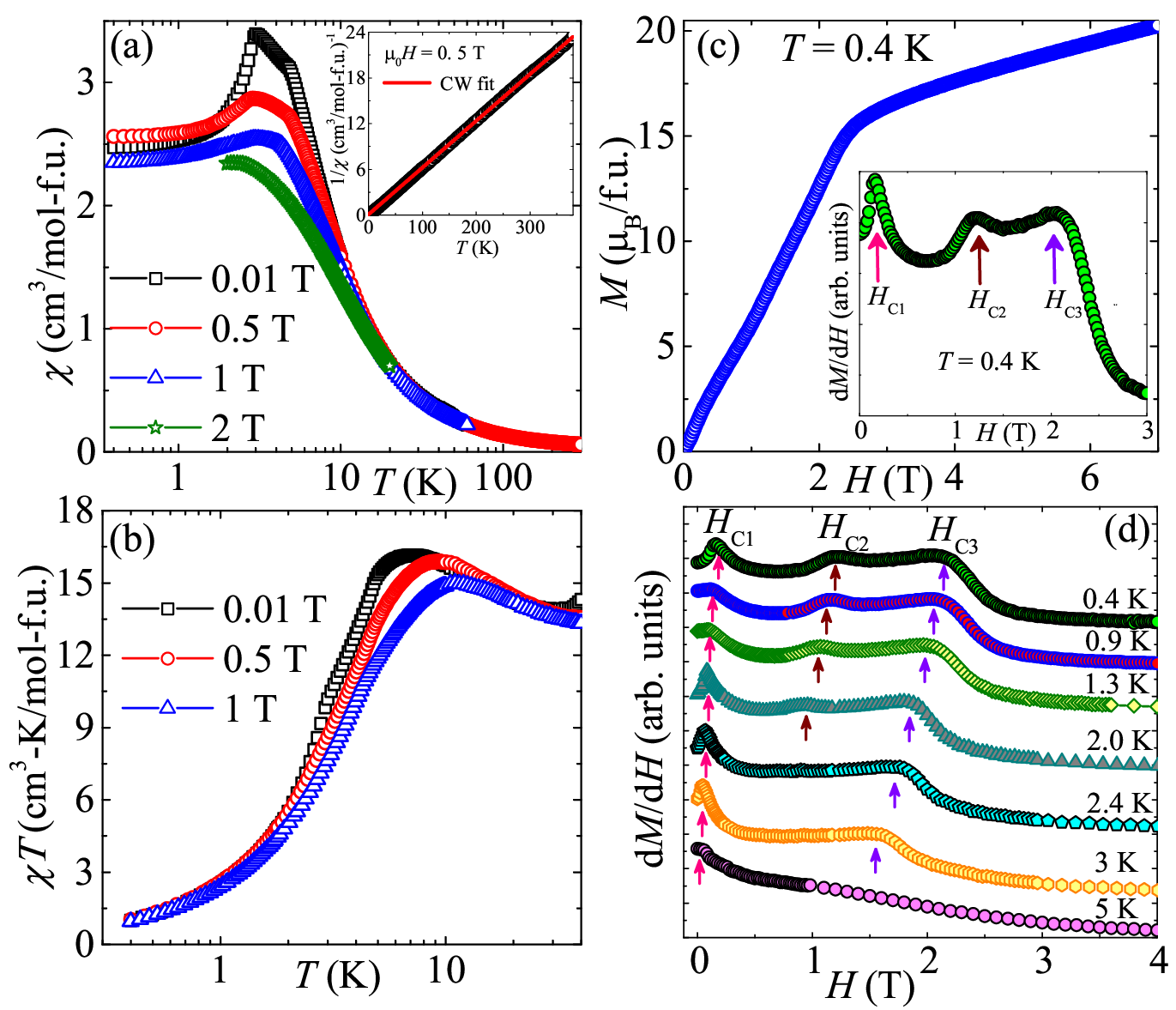}
\caption{\label{Fig3}(a) $\chi$ vs $T$ in different fields. Inset: $1/\chi$ vs $T$ for $\mu_0 H = 0.5$~T with the Curie-Weiss fit. (b) $\chi T$ vs $T$ in different fields. (c) Magnetic isotherm ($M$ vs $H$) and its derivative in the inset at $T=0.4$~K . (d) Derivative of isothermal magnetization vs $H$ in different temperatures showing the field induced transitions marked by $H_{\rm{C1}}$, $H_{\rm{C2}}$, and $H_{\rm{C3}}$.}
\end{figure*}
Magnetization data for MCGO are presented in Fig.~\ref{Fig3}. Temperature-dependent magnetic susceptibility $\chi~[\equiv M/H]$ measured in different applied fields is shown in Fig.~\ref{Fig3}(a). It displays two clear anomalies at $T_{\rm N1} \simeq 4.7$~K and $T_{\rm N2} \simeq 2.8$~K in $\mu_0 H = 0.01$~T, reflecting two successive magnetic transitions~\cite{Mohanty104424}. With increasing field, both the anomalies are suppressed towards low temperatures, typically expected for an AFM ordering. The inverse susceptibility ($1/\chi$) as a function of temperature [inset of Fig.~\ref{Fig3}(a)] exhibits linear behavior in the high-temperature paramagnetic (PM) regime. For a tentative estimation of magnetic parameters, we fitted the data above $80$~K by the modified Curie-Weiss (CW) law
\begin{equation}
\chi(T)=\chi_0+\frac{C}{(T-\theta_{\rm CW})}.
\label{Eq1}
\end{equation}
Here, $\chi_0$ is the temperature-independent susceptibility, $C$ is the CW constant, and $\theta_{\rm CW}$ is the characteristic CW temperature. The fit yields $\chi_0 \simeq -1.79 \times 10^{-5}$~cm$^3$/mol, $C \simeq 16.31$~cm$^3$K/mol, and $\theta_{\rm CW} \simeq -1$~K. From the value of $C$ the effective moment is calculated to be $\mu_{\rm eff} \simeq 11.42~\mu_{\rm B}$. Theoretically, $\mu_{\rm eff}$ for a compound containing two magnetic ions can be calculated as $\mu_{\rm eff}^2 = n_1 \mu_1^2 + n_2 \mu_2^2$ where, $n_1$ and $n_2$ are the number of magnetic ions present in the compound and $\mu_1$ and $\mu_2$ are their respective spin-only effective moments~\cite{Nath224513,Subramanian24}. In the formula unit, MCGO contains $n_1 = 3$ magnetic Mn$^{2+}$ ions with spin $S_1 = 5/2$ and $n_2 = 2$ magnetic Cr$^{3+}$ ions with spin $S_2 = 3/2$. The obtained $\mu_{\rm eff} \simeq 11.42$~$\mu_{\rm B}$ from the CW fit is found to be close to the calculated value of $\mu_{\rm eff} = 11.61$~$\mu_{\rm B}$, considering all these magnetic ions. The small and negative value of $\theta_{\rm CW}$ reflects co-existence of FM and AFM interactions with the dominant one being AFM.


Figure~\ref{Fig3}(b) presents the $\chi T$ vs $T$ plot for different applied fields. As one goes down in temperature, $\chi T$ increases continuously, passes through a maximum around $10$~K, and then falls rapidly towards zero. The initial rise and gradual fall are clear signatures of FM and AFM correlations at high and low-$T$s, respectively~\cite{Savina104447}. Thus, the coexistence of FM and AFM interactions is inferred from the small negative value of $\theta_{\rm CW}$ as well as from the $\chi T$ behavior.


A magnetic isotherm ($M$ vs $H$) measured at $T = 0.4$~K is shown in Fig.~\ref{Fig3}(c). $M$ increases with $H$, shows weak slope changes at several intermediate fields followed by a clear kink around $\sim 2.5$~T. At higher fields (above 2.5~T), $M$ increases linearly but with a much lower slope compared to the initial low-field part. By extrapolating the linear higher-field part to zero field, we determined that the magnetization of 15\,$\mu_B$/f.u. is reached at 2.5\,T which is well below the maximum value of saturation magnetization of the entire spin system, $M_{\rm S} = g(n_1 S_1 + n_2 S_2)\mu_{\rm B} = 21~\mu_{\rm B}$, taking $g=2$, $n_1 = 3$, $S_1 = 5/2$, $n_2 = 2$, and $S_2 = 3/2$. Even at $\mu_0 H = 7$~T, the value of $M \simeq 20.2$~$\mu_{\rm B}$/f.u. is still below the expected saturated magnetization.

The derivative of magnetization with respect to field ($dM/dH$) vs $H$ presented in Fig.~\ref{Fig3}(d) for different temperatures clearly visualizes the slope changes at the critical fields $H_{\rm C1}$, $H_{\rm C2}$, and $H_{\rm C3}$. $dM/dH$ at $T=0.4$~K is shown separately in the inset of Fig.~\ref{Fig3}(c) for a better visualization of these three critical fields. These field-induced features are more pronounced at low temperatures, shift with temperature, and then disappear at high temperatures. 
The appearance of multiple field-induced transitions indicates strong magnetic frustration in the compound~\cite{Deen014419}.

\subsection{Heat Capacity}
\begin{figure*}
\includegraphics[width=\textwidth]{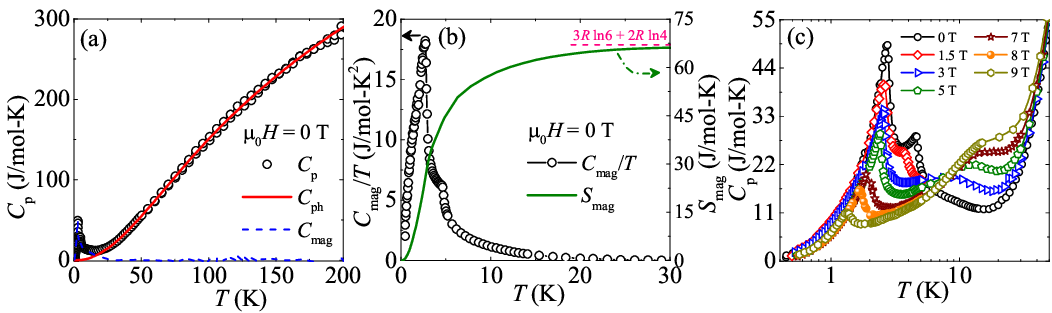}	
\caption{\label{Fig4} (a) $C_{\rm p}$ vs $T$ in zero field. The red solid line represents the phonon contribution ($C_{\rm ph}$), while the blue dashed line indicates the magnetic contribution ($C_{\rm mag}$). (b) $C_{\rm mag}/T$ and $S_{\rm mag}$ vs $T$ in the left and right y-axes, respectively. (c) $C_{\rm p}$ vs $T$ in the low-$T$ regime measured in different fields.}
\end{figure*}
Temperature-dependent heat capacity ($C_{\rm p}$) measured in zero field is shown in Fig.~\ref{Fig4}(a). In a magnetic insulator, the total heat capacity $C_{\rm p}(T)$ is the sum of two major contributions: phonon contribution $C_{\rm ph}(T)$, which dominates in the high-temperature region, and magnetic contribution $C_{\rm mag}(T)$ that dominates in the low-temperature region depending upon the strength of the exchange interactions. In order to extract $C_{\rm mag}(T)$ from $C_{\rm p}(T)$, first $C_{\rm ph}(T)$ was estimated fitting the \mbox{high-$T$} $C_{\rm p}$ data by a linear combination of one Debye [$C_{\rm D}(T)$] and three Einstein [$C_{\rm E}(T)$] terms (Debye-Einstein model) as~\cite{Gopal2012,Magar054076,Sebastian064413}
\begin{equation}
	C_{\rm ph}(T)=f_{\rm D}C_{\rm D}(\theta_{\rm D},T)+\sum_{i = 1}^{3}g_{i}C_{{\rm E}_i}(\theta_{{\rm E}_i},T).
	\label{Eq2}
\end{equation}
The first term in Eq.~\eqref{Eq2} takes into account the acoustic modes, called the Debye term with the coefficient $f_{\rm D}$ and 
\begin{equation}
	C_{\rm D} (\theta_{\rm D}, T)=9nR\left(\frac{T}{\theta_{\rm D}}\right)^{3} \int_0^{\frac{\theta_{\rm D}}{T}}\frac{x^4e^x}{(e^x-1)^2} dx.
	\label{Eq3}
\end{equation}
Here, $x=\frac{\hbar\omega}{k_{\rm B}T}$, $\omega$ is the frequency of oscillation, $R$ is the universal gas constant, and $\theta_{\rm D}$ is the characteristic Debye temperature.
The second term in Eq.~\eqref{Eq2} accounts for the optical modes of the phonon vibration, known as the Einstein term with the coefficient $g_i$ and
\begin{equation}
C_{\rm E}(\theta_{\rm E}, T) = 3nR\left(\frac{\theta_{\rm E}}{T}\right)^2 
\frac{e^{\left(\frac{\theta_{\rm E}}{T}\right)}}{[e^{\left(\frac{\theta_{\rm E}}{T}\right)}-1]^{2}}.
\label{Eq7} 
\end{equation}
Here, $\theta_{\rm E}$ is the characteristic Einstein temperature. The coefficients $f_{\rm D}$, $g_1$, $g_2$, and $g_3$ represent the fraction of atoms that contribute to their respective parts. These values are taken in such a way that their sum should be equal to one. The zero-field $C_{\rm p}(T)$ data above $\sim20$~K are fitted by Eq.~\eqref{Eq2} [red solid line in Fig.~\ref{Fig4}(a)] and the obtained parameters are $f_{\rm D} \simeq 0.06$, $g_1 \simeq 0.18$, $g_2 \simeq 0.37$, $g_3 \simeq 0.39$, $\theta_{\rm D} \simeq 115$~K, $\theta_{{\rm E}_1} \simeq 170$~K, $\theta_{{\rm E}_2} \simeq 360$~K, and $\theta_{{\rm E}_3} \simeq 700$~K. Finally, the high-$T$ fit was extrapolated down to low temperatures and $C_{\rm mag}(T)$ [blue dashed line in Fig.~\ref{Fig4}(a)] was estimated by subtracting $C_{\rm {ph}}(T)$ from $C_{\rm p}(T)$. Figure~\ref{Fig4}(b) presents $C_{\rm mag}(T)/T$ and the corresponding magnetic entropy [$S_{\rm{mag}}(T) = \int_{\rm 0.4\,K}^{T}\frac{C_{\rm {mag}}(T')}{T'}dT'$]. The obtained magnetic entropy, which saturates above 20~K, approaches a value $S_{\rm mag} \simeq 67.15$~J/mol-K. This value is close to the expected theoretical values of $S_{\rm mag} = n_1 \times R \ln(2 S_1 + 1)+ n_2 \times R \ln(2 S_2 + 1) = 67.74$~J/mol-K.

At low temperatures, zero-field $C_{\rm p}(T)$ shows two well-defined anomalies at $T_{\rm N1} \simeq 4.5$~K and $T_{\rm N2} \simeq 2.7$~K, confirming two successive magnetic transitions. To gain more information about the magnetic transitions, we measured $C_{\rm p}(T)$ in different applied fields [see Fig.~\ref{Fig4}(c)]. With increasing field, the height of the peaks is reduced substantially and the peak position shifts towards low temperatures, as typical of AFM transitions. For $\mu_0H > 1.5$~T, $T_{\rm N1}$ disappears completely from the measurement window while $T_{\rm N2}$ is shifted to 1.34~K for $H=9$~T. 
Concurrently, another broader maximum emerges above $\sim 1.5$~T and is driven higher in temperatures with increasing field. This is likely due to the redistribution of entropy where the entropy, which was released at the magnetic transition in low fields, is shifted towards higher temperatures as the field is increased.

\subsection{Neutron Diffraction}
\begin{figure}
\includegraphics[width=\columnwidth]{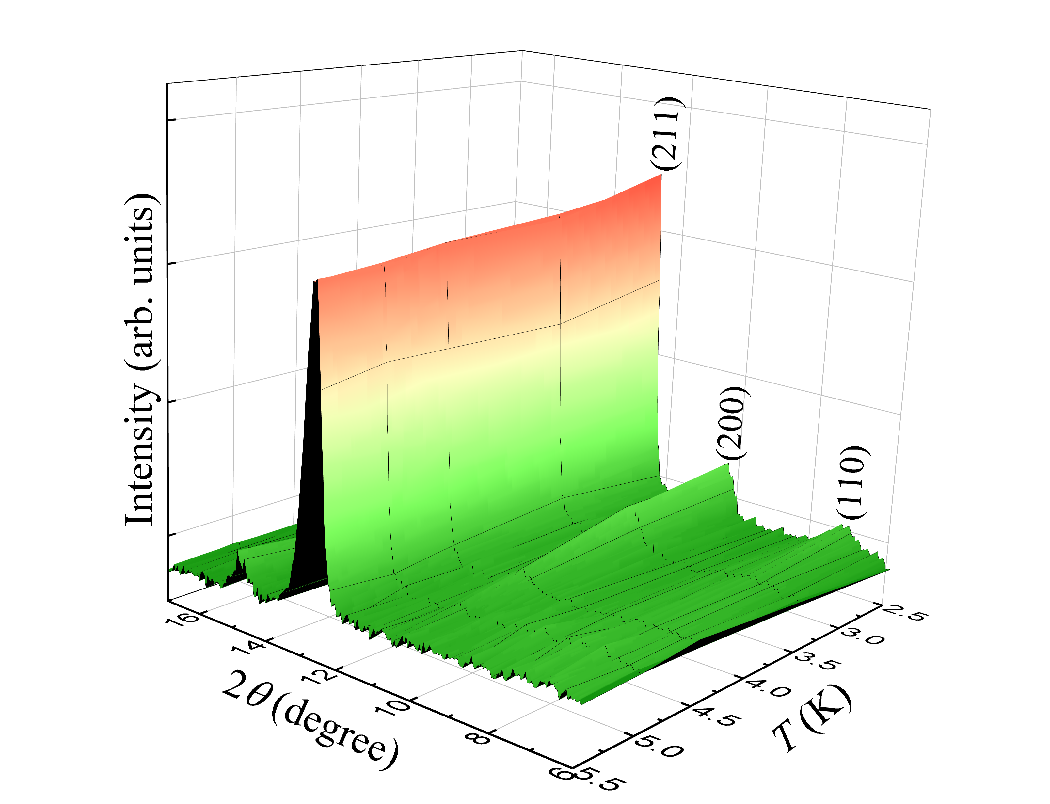}	
\caption{\label{Fig5} 3D representation of the NPD data collected around the transitions. Only the growth of magnetic reflections are pinpointed.}
\end{figure}
\begin{figure}
\includegraphics[width=\columnwidth]{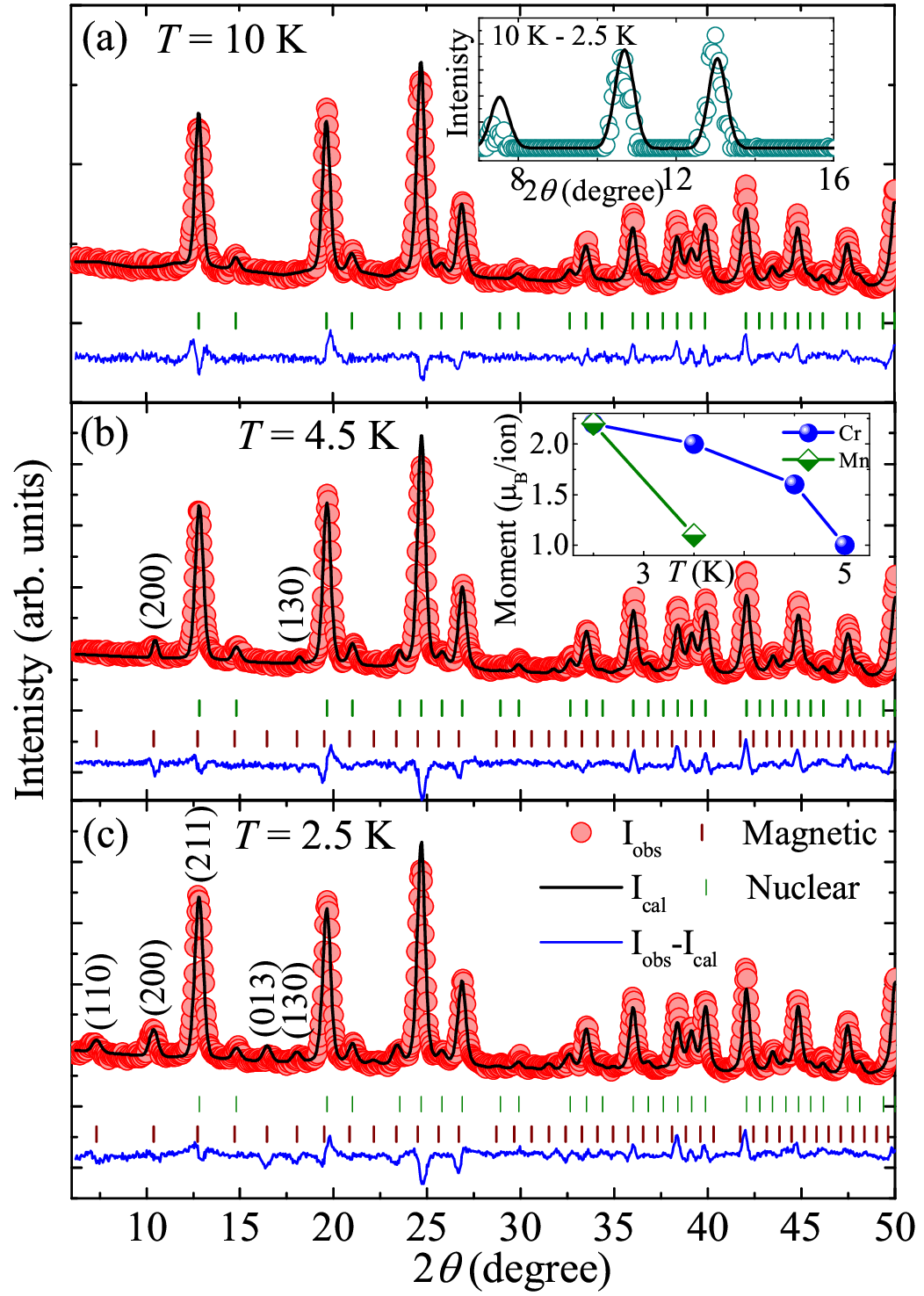}	
\caption{\label{Fig6} The NPD data presented at three different temperatures: (a) well above the magnetic transitions (10~K) with only nuclear peaks, (b) below $T_{\rm N1}$ with two additional magnetic peaks, and (c) below both $T_{\rm N1}$ and $T_{\rm N2}$, clearly showing five magnetic peaks. Rietveld fits are shown as solid black lines. Vertical bars are the allowed nuclear (top row) and magnetic (bottom row) Bragg peaks. Only the magnetic peaks in the data are indexed. Inset of (a): the difference in data between 2.5~K and 10~K highlighting only the magnetic peaks and the solid line is the Rietveld fit to the data. Inset of (b): ordered moments of Cr$^{3+}$ and Mn$^{2+}$ vs temperature.}
\end{figure}
In order to resolve the magnetic structure, neutron powder diffraction (NPD) data were collected at temperatures between 2.5~K to 50~K. Figure~\ref{Fig5} presents the 3D view of the temperature evolution of the low-angle peaks. At high temperatures, all the peaks are found to be arising from the nuclear reflections. At low temperatures ($T<T_{\rm N1}$), several extra peaks with low intensities appear, indicating the formation of LRO. For $T < T_{\rm N1}$, two magnetic Bragg peaks are detected at $2\theta \sim 17.98\degree$ and $\sim 10.37\degree$, while for $T < T_{\rm N2}$ two more peaks are observed at $\sim 16.48\degree$ and $\sim 7.32\degree$ and the intensity of the nuclear peak at $\sim 12.78\degree$ is found to be enhanced. Due to the low intensity, some of the magnetic peaks are not clearly visible in the 3D plot. The individual plots in Fig.~\ref{Fig6} highlight these peaks. The appearance of the distinct magnetic Bragg peaks below each of the $T_{\rm N1}$ and $T_{\rm N2}$ confirms that MCGO undergoes two AFM transitions.

Rietveld refinement is performed at three different temperatures, well above the magnetic transitions (10~K), below $T_{\rm N1}$ (4.5~K), and below $T_{\rm N2}$ (2.5~K) (see Fig.~\ref{Fig6}). All the peaks in the NPD data at 10~K are nuclear in origin and well fitted by the cubic crystal structure with the $Ia\bar{3}d$ space group [Fig.~\ref{Fig6}(a)]. 

As shown in Fig.~\ref{Fig6}(b), the data collected at 4.5~K (i.e. below $T_{\rm N1}$) shows two new peaks. Both the magnetic peaks could be indexed using the propagation vector $k=(0,0,0)$ and space group $I-1$. The symmetry analysis shows that these magnetic reflections can be modeled by taking the collinear AFM order within the Cr$^{3+}$ sublattice where individual moments are aligned along the $[1,0,0]$ direction. The magnetic peaks are identified to be $(200)$ and $(130)$. The refined magnetic structure of the Cr$^{3+}$ sublattice is depicted in Fig.~\ref{Fig1}(b) in which the Cr$^{3+}$ moments are aligned parallel to each other (FM) in the $ac$-plane and antiparallel (AFM) along the $b$-direction. This confirms that below $T_{\rm N1}$, Cr$^{3+}$ sublattice is ordered in a collinear AFM fashion. Further, all the magnetic peaks in the NPD data at 2.5~K [Fig.~\ref{Fig6}(c)] could be indexed by the same propagation vector and space group by introducing the magnetic moment of the Mn$^{2+}$ sublattice.
The magnetic peaks could be indexed as $(110)$ and $(013)$, while the nuclear peak with the enhanced intensity is identified as $(211)$. Therefore, we conclude that $T_{\rm N2}$ is due to the ordering of the Mn$^{2+}$ sublattice. The ordering of the Mn$^{2+}$ sublattice is non-collinear AFM type as shown in Fig.~\ref{Fig1}(c). The magnetic structure of the Mn$^{2+}$ sublattice is in close agreement with the previous report~\cite{Golosovskii461,Gukasov2869}. Moreover, we also performed Rietveld refinement of only the magnetic reflections obtained by taking a difference of the 2.5~K and 10~K data as shown in the inset of Fig.~\ref{Fig6}(a). The magnetic moment values obtained from the refinement are consistent with the values obtained from the full dataset refinement, confirming good quality of the fit.

The ordered moments for both the magnetic ions are plotted as a function of temperature in the inset of Fig.~\ref{Fig6}(b). At $T =2.5$~K, the refined value of the ordered moment of Cr$^{3+}$ is $\mu \simeq 2.2$~$\mu_{\rm B}$ and that of Mn$^{2+}$ is $\mu \simeq 2.2$~$\mu_{\rm B}$. These values are considerably reduced compared to the expected spin-only values of $\sim 3$~$\mu_{\rm B}$ for Cr$^{3+}$ and $\sim 5$~$\mu_{\rm B}$ for Mn$^{2+}$, respectively. The reduction in the magnetic moment from its classical value is commonly observed in frustrated magnets~\cite{Islam134433,Sebastian104425}, although in the present case thermal fluctuations may also play a role because the measurement temperature of 2.5~K is close to $T_{\rm N2}\simeq2.7$~K and more than half of $T_{\rm N1}\simeq 4.5$~K.

\begin{figure}
\includegraphics[width=\columnwidth]{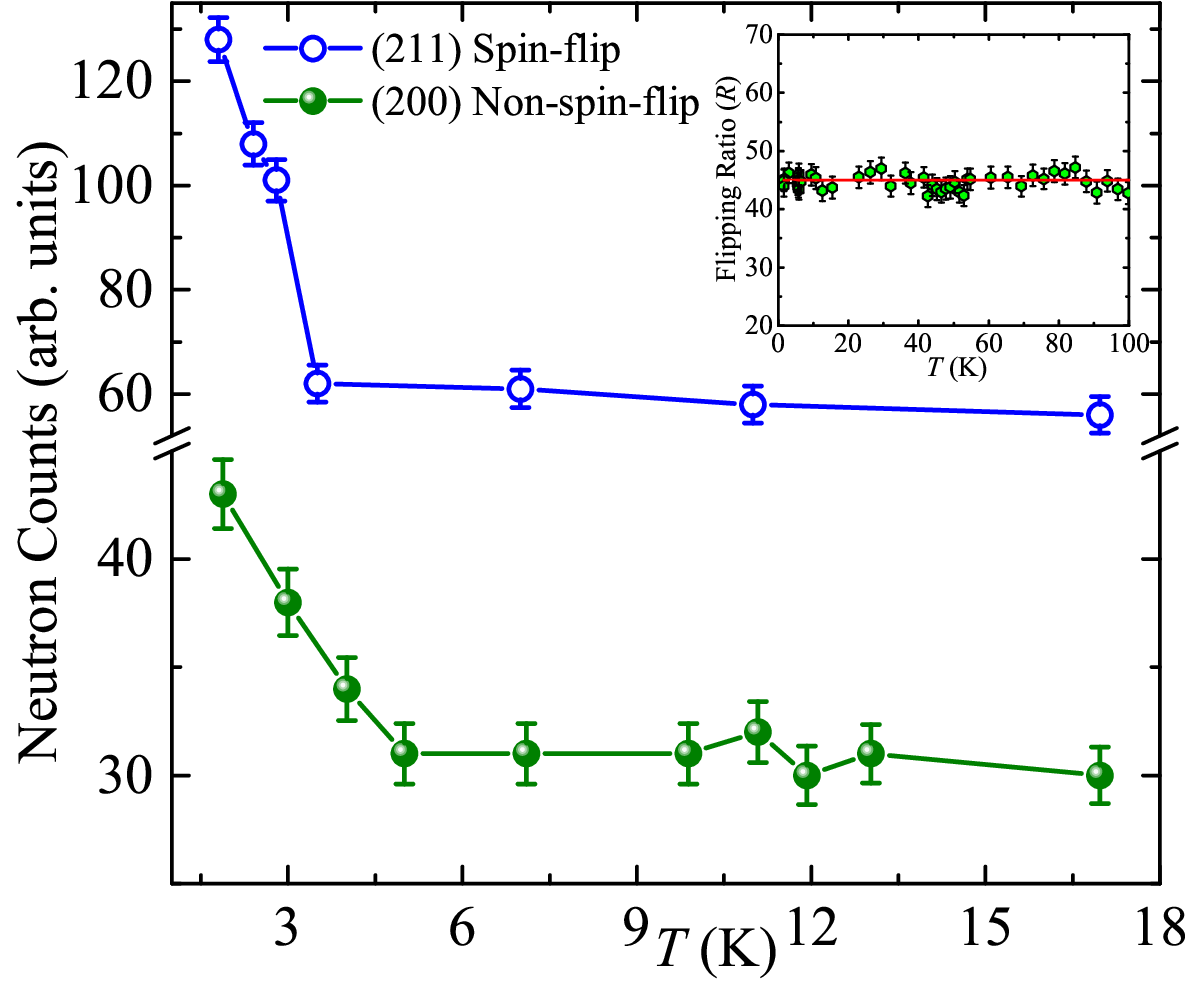}	
\caption{\label{Fig7} Temperature dependence of the magnetic reflections (200) and (211) in the non-spin-flip and spin-flip channels, corresponding to the Cr$^{3+}$ and Mn$^{2+}$ sublattices. Inset: temperature variation of the flipping ratio ($R$) down to $1.7$~K.}
\end{figure}
To investigate a possible presence of FM correlations, we have performed one-dimensional (1D) neutron-depolarization measurements during which the polarization analysis of both the incident and diffracted neutron beams was performed. In these experiments, the rotation of the neutron polarization vector after transmission through the sample provides direct information about the presence and characteristics of FM correlations within the material, over a length scale of $100-1000$~\AA. The inset of Fig.~\ref{Fig7} depicts temperature variation of the flipping ratio ($R$) down to $1.7$~K, where no depolarization of the neutron beam is observed, indicating the absence of FM or ferrimagnetic domains/clusters of the mentioned length scale under an applied field of $50$~Oe. However, clusters of smaller length scales cannot be completely ruled out even though our diffraction data at low angles do not show any broad features of diffuse scattering that could be indicative of a short-range magnetic order.

For a better understanding of the complex magnetic structure, we used the polarized neutron spectrometer to measure temperature variation in the peak intensity of the magnetic reflections (200) and (211) in both the non-spin-flip and spin-flip channels. In this spectrometer, the neutron polarization direction is perpendicular to the scattering vector ($P\perp Q$ geometry). As we have mentioned earlier, the Bragg reflection (200) is purely magnetic and represents the ordering of the Cr$^{3+}$ sublattice whereas the enhanced intensity of the (211) nuclear peak represents the magnetic contribution from the Mn$^{2+}$ sublattice. The magnetic contribution of the (211) peak was confirmed by measuring the neutron intensity in the spin-flip channel. As presented in Fig.~\ref{Fig7}, the neutron counts corresponding to (200) and (211) increase abruptly below $T_{\rm N1}$ and $T_{\rm N2}$, respectively, thus further supporting the independent ordering of the Cr$^{3+}$ and Mn$^{2+}$ sublattices in zero field.

DFT calculations reveal AFM nearest-neighbor exchange couplings within both Cr$^{3+}$ and Mn$^{2+}$ sublattices. We obtain $J_{\rm Mn-Mn}=3.0$\,K as well as $J_{\rm Cr-Cr}=1.9$\,K and $J_{\rm Cr-Cr}'=2.8$\,K where the former and latter values stand for the Cr--Cr contacts with and without the GeO$_4$ bridge, respectively. The nearest-neighbor interaction between the sublattices is FM in nature, $J_{\rm Mn-Cr}=-2.8$\,K. The interactions beyond nearest neighbors do not exceed 0.5\,K and can be neglected within the minimum microscopic model.

Magnetic ground state of MCGO can be inferred from the AFM intra-sublattice interactions. Indeed, the nearest-neighbor couplings $J_{\rm Cr-Cr}$ and $J_{\rm Cr-Cr}'$ form a non-frustrated 8-coordinated (bcc-like) Cr$^{3+}$ sublattice that develops the collinear AFM order (Fig.~\ref{Fig1}b). By contrast, the Mn$^{2+}$ sublattice comprises triangles and adopts a non-coplanar configuration with the $120^{\circ}$-like arrangement of spins on each of the triangles (Fig.~\ref{Fig1}c). This state belongs to the manifold of the classically degenerate states of the hyperkagome lattice~\cite{Hopkinson037201}. It is remarkable that the Mn$^{2+}$ sublattice orders at a lower temperature compared to its Cr$^{3+}$ counterpart, despite the larger spin of Mn and the stronger magnetic couplings, $J_{\rm Mn-Mn}> J_{\rm Cr-Cr}'$. The lower ordering temperature of the Mn$^{2+}$ sublattice can be traced back to its frustrated nature and to the lower coordination number (4 for Mn vs 8 for Cr).

Whereas Cr--Cr interactions are long-range in nature, the Mn--Mn and Mn--Cr interactions involve superexchange via one oxygen atom and can be analyzed in terms of the Goodenough-Kanamori-Anderson rules~\cite{Kanamori87}. The bond angles are quite similar, $102.4^{\circ}$ for $J_{\rm Mn-Mn}>0$ vs. $99.0^{\circ}$ and $103.5^{\circ}$ for $J_{\rm Mn-Cr}<0$. The different signs of these couplings should be then ascribed to the different electronic configurations of Mn$^{2+}$ and Cr$^{3+}$. Whereas all five $d$-orbitals of Mn$^{2+}$ are half-filled, two of the Cr$^{3+}$ $d$-orbitals are empty. Hoppings between the half-filled and empty orbitals of Mn$^{2+}$ and Cr$^{3+}$, respectively, give rise to a FM contribution that appears to be dominant in $J_{\rm Mn-Cr}$. In the experimental magnetic structure, the contribution of $J_{\rm Mn-Cr}$ vanishes because each Cr$^{3+}$ spin is coupled to two oppositely aligned Mn$^{2+}$ spins [Fig.~\ref{Fig1}(d)]. The FM coupling $J_{\rm Mn-Cr}$ is incompatible with the non-coplanar order of the Mn$^{2+}$ sublattice. It is then natural that the two sublattices order independently from each other at two distinct AFM transitions.

\subsection{Phase Diagram}
\begin{figure}
\includegraphics[width=\columnwidth]{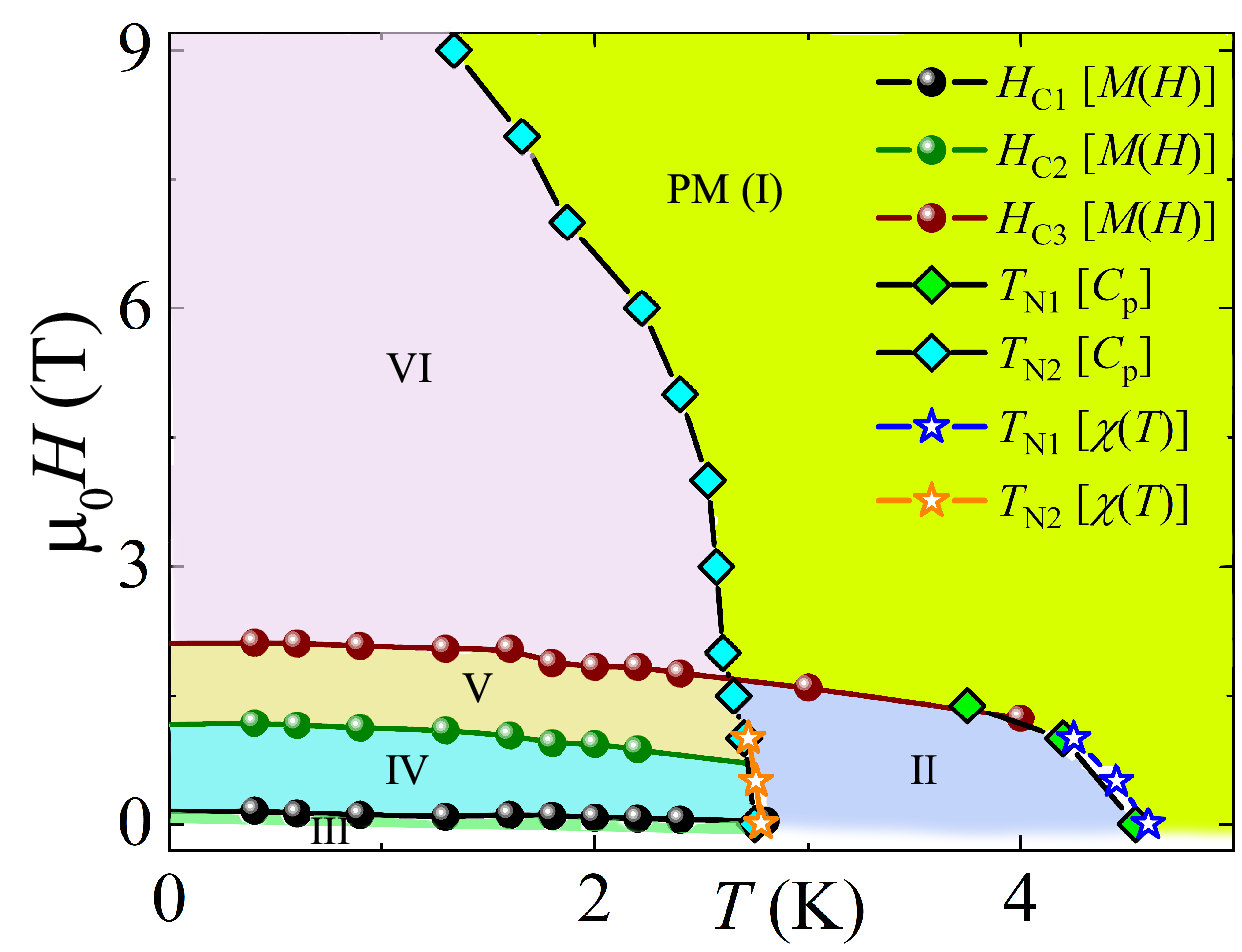}	
\caption{\label{Fig8}$H-T$ phase diagram obtained from the magnetic isotherm, heat capacity, and susceptibility data.}
\end{figure}
The values of $T_{\rm N1}$ and $T_{\rm N2}$ obtained from $\chi(T)$ and $C_{\rm p}(T)$ along with $H_{\rm C1}$, $H_{\rm C2}$, and $H_{\rm C3}$ obtained from the magnetic isotherms are summarized in Fig.~\ref{Fig8}. The $H-T$ phase diagram features six distinct phases. Phases II and III represent the zero-field ordered states of the Cr$^{3+}$ and Mn$^{2+}$ sublattices, respectively. When magnetic field is applied, three new phases (IV, V, and VI) emerge.
This complex phase diagram can be attributed to the strongly frustrated nature of the spin lattice in the garnet structure. Single-crystal neutron scattering experiments in magnetic fields would be necessary to unveil the precise nature of these phases. Similar type of complex phase diagrams is commonly found in other garnet compounds due to their underlying frustration~\cite{Florea220413,Deen014419}.

\subsection{Magnetocaloric Effect}
\begin{figure*}
\includegraphics[width=\textwidth]{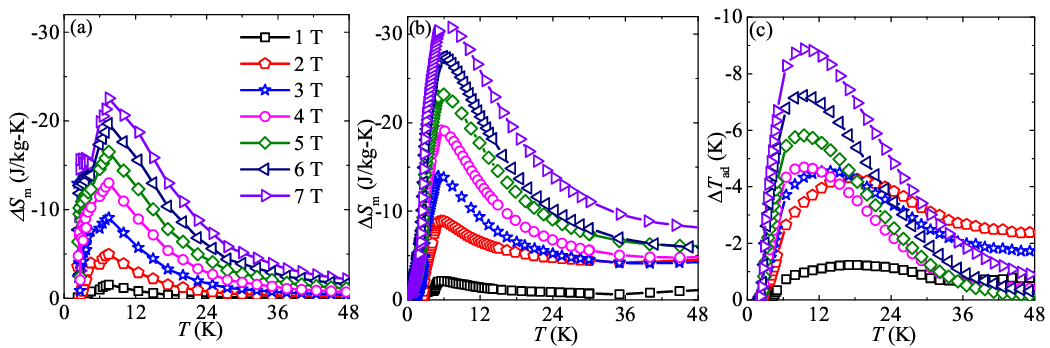} 
\caption{\label{Fig9}(a) Entropy change ($\Delta S_{\rm m}$) as a function of $T$ calculated for the field changes of $\Delta H = 1$~T to $7$~T using the magnetization data. (b) $\Delta S_{\rm m}$ vs $T$ calculated for the same $\Delta H$ values using field-dependent heat capacity data. (c) Adiabatic temperature change ($\Delta T_{\rm ad}$) vs $T$ calculated for the same $\Delta H$ values using heat capacity data.}
\end{figure*}
\begin{figure}
\includegraphics[width=\columnwidth]{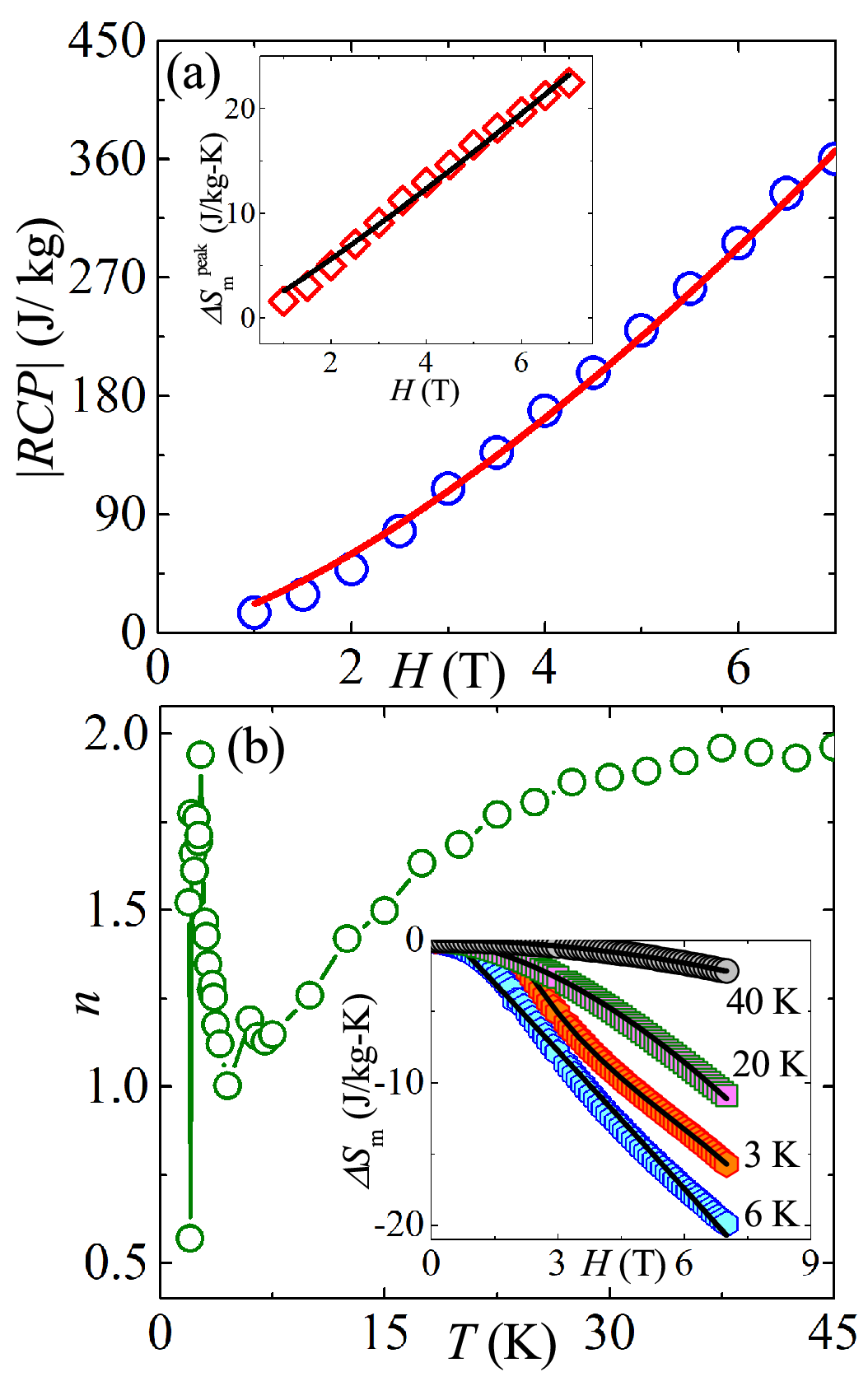} 
\caption{\label{Fig10} (a) Relative cooling power ($RCP$) as a function of field ($H$) obtained from Fig.~\ref{Fig9}(a). Inset: peak position $\Delta S_{\rm m}^{\rm peak}$ as a function of $H$ taken from Fig.~\ref{Fig9}(a). (b) Temperature dependence of the power-law exponent $n$. Inset: $\Delta S_{\rm m}$ vs $H$ curves at different temperatures, with the solid lines representing the power-law fits.}
\end{figure}
Temperatures in the sub-Kelvin range can be attained by employing the MCE~\cite{Pecharsky44}. In this process, magnetic field is applied to the material isothermally and then removed adiabatically. Therefore, MCE can be quantified by the isothermal entropy change ($\Delta S_{\rm m}$) and adiabatic temperature change ($\Delta T_{\rm ad}$) with respect to the change in the applied field ($\Delta H$). MCGO features two magnetic ions with large magnetic moments ($S = 5/2$ for Mn$^{2+}$ and $S = 3/2$ for Cr$^{3+}$) and the double magnetic transition. Therefore, MCGO is expected to exhibit large $\Delta S_{\rm m}$ and the associated cooling power at low temperatures. $\Delta S_{\rm m}$ can be calculated from magnetization isotherms ($M$ vs $H$) measured in close temperature steps around the transitions. Using Maxwell's thermodynamic relation, $(\partial S/\partial H)_T = (\partial M/\partial T)_H$, we estimate $\Delta S_{\rm m}$ as~\cite{Islam134433,Singh111941}
\begin{equation}
\Delta S_{\rm m} (H, T) =\int_{H_{\rm i}}^{H_{\rm f}}\dfrac{dM}{dT}dH.
\label{Sm_magnetization}
\end{equation}
Figure~\ref{Fig9}(a) displays the plot of $\Delta S_{\rm m}$ as a function of temperature for different $\Delta H$ values, calculated using Eq.~\eqref{Sm_magnetization}. It features caret-like shape with its maximum centered around $T_{\rm N}$, typically expected for materials with the second-order magnetic transition. A large MCE characterized by the maximum entropy change of $\Delta S_{\rm m} \simeq -23$~J/kg-K is obtained for the field change of 7~T.

In order to cross-check the large value of $\Delta S_{\rm m}$, we also estimated $\Delta S_{\rm m}$ from the heat capacity data measured in zero field and at higher fields up to 7~T. First, we calculate the total entropy at a given field as
\begin{equation}
	S(T)_H = \int_{T_{i}}^{T_{f}} \frac{ C_{\rm p}(T)_{H}}{T} dT,
	\label{Sm_heatcapacity}
\end{equation}
where $C_{\rm p}(T)_H$ is the heat capacity at field $H$, whereas $T_{i}$ and $T_{f}$ are the initial and final temperatures, respectively. Next, we calculate $\Delta S_{\rm m}$ by taking the difference of the total entropy at non-zero and zero fields as $\Delta S_{\rm m}(T)_{\Delta H} = [S(T)_{H} - S(T)_{0}]_T$. Here, $S(T)_{H}$ and $S(T)_{0}$ are the total entropy in the presence of $H$ and in zero field, respectively. Figure~\ref{Fig9}(b) presents the estimated $\Delta S_{\rm m}$ as a function of $T$ for different $\Delta H$ values. The shape and peak position of the curves are nearly identical to the curves obtained from the magnetic isotherms in Fig.~\ref{Fig9}(a) but with a slightly enhanced value of $\Delta S_{\rm m}$~\cite{Pecharsky565}.

The adiabatic temperature change $\Delta T_{\rm ad}$ can be estimated from either the combination of zero-field heat capacity and the magnetic isotherm data or the heat capacity alone measured in different magnetic fields. The estimation of $\Delta T_{\rm ad}$ following the former method may not always give reliable results, as discussed in Ref.~\cite{Magar054076}. Therefore, we estimated $\Delta T_{\rm ad}$ from the heat capacity data alone by taking the difference in temperatures corresponding to two different fields with same entropy value as
\begin{equation}
	\Delta T_{\rm ad}(T)_{\Delta H} = [T(S)_{H_f} - T(S)_{H_i}].
	\label{proper}
\end{equation}
$\Delta T_{\rm ad}$ vs $T$ for $1 \leq \Delta H \leq 7$~T calculated by this method is shown in Fig.~\ref{Fig9}(c). The maximum value at the peak position is found to be $\Delta T_{\rm ad} \simeq 9$~K for $\Delta H = 7$~T.

Another important parameter that determines the cooling performance of the material is the relative cooling power ($RCP$). $RCP$ is a measure of the amount of heat transferred between the cold and hot reservoirs in a refrigeration cycle. Mathematically, it can be expressed as
\begin{equation}
	{RCP} = \int_{T_{\rm {cold}}}^{T_{\rm hot}} \Delta S_{\rm m}(T,H)~dT,
\end{equation}
where $T_{\rm cold}$ and $T_{\rm hot}$ correspond to the temperatures of the cold and hot reservoirs, respectively. The $RCP$ can be approximated as
\begin{equation}
	|{ RCP}|_{\rm approx} = \Delta S_{\rm m}^{\rm peak} \times \delta T_{\rm FWHM},
\end{equation}
where $\Delta S_m^{\rm peak}$ and $ \delta T_{\rm FWHM}$ are the maximum value of the entropy change and full-width at half-maximum of the $\Delta S_m$ vs $T$ curves, respectively. $RCP$ as a function of $H$ for MCGO calculated using the $\Delta S_m$ data from Fig.~\ref{Fig9}(a) is plotted in Fig.~\ref{Fig10}(a). The maximum value of $RCP$ is about $\sim 360$~J/kg for 7~T. 

Magnetic cooling is a cyclic process involving repeated demagnetization of the material. Materials with first-order phase transitions are undesirable for the cyclic operation because of the energy loss via magnetic or thermal hysteresis~\cite{Franco305}. Materials with second-order phase transitions are better suited for the commercial use. To characterize the nature of the phase transition in a given material, one can further analyze the field dependence of $RCP$ and $\Delta S_m^{\rm peak}$ as shown in Fig.~\ref{Fig10}(a)~\cite{Singh6981}. We fitted the $RCP(H)$ and $\Delta S_m^{\rm peak}(H)$ data by power laws of the form $RCP \propto H^{N}$ and $|\Delta S_m^{\rm peak}| \propto H^n$, respectively. The values of the exponents are estimated to be $N \simeq 1.44$ and $n \simeq 1.12$. These exponents are related to the critical exponents $\beta$, $\gamma$, and $\delta$ as $N=1+(1/\delta)$ and $n = 1+\frac{\beta-1}{\beta+\gamma}$. Using the value of $N$ and $n$ in the above relations along with the Widom formula [$\delta=1+(\gamma/\beta)]$ yields $\beta=1.37$, $\gamma=1.75$, and $\delta=2.27$. These values of the critical exponents do not fall under any known universality class~\cite{Islam134433}.

Temperature dependence of $n$ can be used to assess the nature of a phase transition. Generally, for a second-order phase transition, the exponent should have the value $n \simeq 2$ in the paramagnetic region ($T >> T_{\rm N}$) and $n \simeq 1$ well below $T_{\rm N}$, while at $T = T_{\rm N}$ it depends on the critical exponents~\cite{Law2680}. On the other hand, for a first-order phase transition, $n$ will take a value much greater than 2~\cite{Law2680}. To obtain the variation of $n$ with temperature, we fitted $\Delta S_{\rm m}$ vs $H$ curves at various temperatures across the transitions using the power law $\Delta S_m \propto H^n$ [see inset of Fig.~\ref{Fig10}(b)]. The obtained $n$ vs $T$ data is plotted in the main panel of Fig.~\ref{Fig10}(b). In the entire measured temperature range, the value of $n$ remains below 2 with a minimum at around the transition temperatures. This confirms second-order nature of the transitions in MCGO and renders it a suitable material for cyclic operations~\cite{Singh6981}.
\begin{table}[h]
	\caption{MCE performance parameters $\Delta T_{\rm ad}$, $\Delta S_m^{\rm peak}$, and $RCP$ for MCGO and representative cooling materials with large magnetic moments. The compound Li$_9$Cr$_3$(P$_2$O$_7$)$_3$(PO$_4$)$_2$ is abbreviated as LCPP.}
	\begin{tabular}{ c c c c c c c c c c c c c}
		\hline\hline
		System & $T_{\rm C}$ or $T_{\rm N}$ & $|\Delta T_{\rm ad}|$ & $|\Delta S_{\rm m}^{\rm peak}|$ & $RCP$ & $\Delta H$ & Refs. \\
		& (K)   & (K)                 & (J/kg-K)            & (J/kg) & (T) & \\
		\hline
		MCGO & 4.5, 2.7 & 9 & 23 & 360 & 7 & This work \\
		HoMnO$_3$ & 5 & 6.5 & 13.1 & 320 & 7 & \cite{Midya142514}\\
		ErMn$_2$Si$_2$ & 4.5 & 12.9 &	25.2 & 365 & 5 & \cite{Lingwei152403}\\	
		EdDy$_2$O$_4$ &	5 & 16 & 25 & 415 & 8 & \cite{Midya132415}\\
		EuHo$_2$O$_4$ &	5 &	12.7 & 30 &	540	& 8 & \cite{Midya132415}\\
		EuTiO$_{3}$ & 5.6 & 21 & 49 & 500 & 7 & \cite{Midya094422}\\
  LCPP & 2.6 & 9 & 31 & 284 & 7 & \cite{Magar054076}\\
		\hline\hline
	\end{tabular}
	\label{MCETable}
\end{table}

We compared the MCE parameters of MCGO with those of the previously studied MCE materials that have their magnetic transitions in the same temperature range (Table~\ref{MCETable}). This comparison suggests that MCGO is on par with other materials and could be used in cryogenic applications. Thus, the magnetic frustration and large spin values are the key factors for reaching large values of $\Delta S_{\rm m}$. Because of frustration, there is a distribution of the magnetic entropy over a wide temperature range, resulting in large $RCP$ values.

\section{Summary}
In summary, we reported the magnetic and magnetocaloric properties of the garnet compound MCGO. It contains two magnetic sublattices: Mn$^{2+}$ ($S=5/2$) forms a geometrically frustrated hyperkagome lattice, whereas Cr$^{3+}$ ($S=3/2$) forms a non-frustrated eight-coordinated bcc-like lattice. MCGO undergoes two consecutive magnetic transitions at $T_{\rm N1}\sim 4.5$~K and $T_{\rm N2}\sim 2.7$~K, as revealed by the magnetization and heat capacity data. The NPD experiments confirm that the two sublattices order independently. The Cr$^{3+}$ sub-lattice undergoes collinear AFM ordering below $T_{\rm N1}$, while the Mn$^{2+}$ sublattice develops a non-coplanar AFM ordering below $T_{\rm N2}$. Both types of order arise from the nearest-neighbor AFM intra-sublattice interactions. The interaction between the sublattices is FM in nature and cancels out in the experimental magnetic structure owing to the frustrated nature of the garnet structure. 
Multiple field-induced transitions are observed in the $M$ vs $H$ curves below $T_{\rm N2}$, giving rise to a complex $H-T$ phase diagram. 
A large MCE characterized by $\Delta S_{\rm m} \sim -23$~J/kg-K, $\Delta T_{\rm ad} \sim 9$~K, and $RCP \sim 360$~J/kg is obtained for the field change of $7$~T, which can be ascribed to the strong frustration in the spin system. This renders MCGO a promising MCE material to achieve low temperatures by adiabatic demagnetization.

\acknowledgments
For financial support, we would like to acknowledge SERB, India bearing sanction Grant No.~CRG/2022/000997 and DRDO, India bearing CARS Grant No.~DLJ/TC/1025/I/73.


\begin{thebibliography}{54}%
	\makeatletter
	\providecommand \@ifxundefined [1]{%
		\@ifx{#1\undefined}
	}%
	\providecommand \@ifnum [1]{%
		\ifnum #1\expandafter \@firstoftwo
		\else \expandafter \@secondoftwo
		\fi
	}%
	\providecommand \@ifx [1]{%
		\ifx #1\expandafter \@firstoftwo
		\else \expandafter \@secondoftwo
		\fi
	}%
	\providecommand \natexlab [1]{#1}%
	\providecommand \enquote  [1]{``#1''}%
	\providecommand \bibnamefont  [1]{#1}%
	\providecommand \bibfnamefont [1]{#1}%
	\providecommand \citenamefont [1]{#1}%
	\providecommand \href@noop [0]{\@secondoftwo}%
	\providecommand \href [0]{\begingroup \@sanitize@url \@href}%
	\providecommand \@href[1]{\@@startlink{#1}\@@href}%
	\providecommand \@@href[1]{\endgroup#1\@@endlink}%
	\providecommand \@sanitize@url [0]{\catcode `\\12\catcode `\$12\catcode
		`\&12\catcode `\#12\catcode `\^12\catcode `\_12\catcode `\%12\relax}%
	\providecommand \@@startlink[1]{}%
	\providecommand \@@endlink[0]{}%
	\providecommand \url  [0]{\begingroup\@sanitize@url \@url }%
	\providecommand \@url [1]{\endgroup\@href {#1}{\urlprefix }}%
	\providecommand \urlprefix  [0]{URL }%
	\providecommand \Eprint [0]{\href }%
	\providecommand \doibase [0]{https://doi.org/}%
	\providecommand \selectlanguage [0]{\@gobble}%
	\providecommand \bibinfo  [0]{\@secondoftwo}%
	\providecommand \bibfield  [0]{\@secondoftwo}%
	\providecommand \translation [1]{[#1]}%
	\providecommand \BibitemOpen [0]{}%
	\providecommand \bibitemStop [0]{}%
	\providecommand \bibitemNoStop [0]{.\EOS\space}%
	\providecommand \EOS [0]{\spacefactor3000\relax}%
	\providecommand \BibitemShut  [1]{\csname bibitem#1\endcsname}%
	\let\auto@bib@innerbib\@empty
	\bibitem [{\citenamefont {Starykh}(2015)}]{Starykh052502}%
	\BibitemOpen
	\bibfield  {author} {\bibinfo {author} {\bibfnamefont {O.~A.}\ \bibnamefont
			{Starykh}},\ }\bibfield  {title} {\bibinfo {title} {{Unusual ordered phases
				of highly frustrated magnets: a review}},\ }\href
	{https://doi.org/10.1088/0034-4885/78/5/052502} {\bibfield  {journal}
		{\bibinfo  {journal} {Rep. Prog. Phys.}\ }\textbf {\bibinfo {volume} {78}},\
		\bibinfo {pages} {052502} (\bibinfo {year} {2015})}\BibitemShut {NoStop}%
	\bibitem [{\citenamefont {Ramirez}(1994)}]{Ramirez453}%
	\BibitemOpen
	\bibfield  {author} {\bibinfo {author} {\bibfnamefont {A.~P.}\ \bibnamefont
			{Ramirez}},\ }\bibfield  {title} {\bibinfo {title} {Strongly geometrically
			frustrated magnets},\ }\href
	{https://doi.org/10.1146/annurev.ms.24.080194.002321} {\bibfield  {journal}
		{\bibinfo  {journal} {Annu. Rev. Mater. Sci.}\ }\textbf {\bibinfo {volume}
			{24}},\ \bibinfo {pages} {453} (\bibinfo {year} {1994})}\BibitemShut
	{NoStop}%
	\bibitem [{\citenamefont {Gardner}\ \emph {et~al.}(2010)\citenamefont
		{Gardner}, \citenamefont {Gingras},\ and\ \citenamefont
		{Greedan}}]{Gardner53}%
	\BibitemOpen
	\bibfield  {author} {\bibinfo {author} {\bibfnamefont {J.~S.}\ \bibnamefont
			{Gardner}}, \bibinfo {author} {\bibfnamefont {M.~J.~P.}\ \bibnamefont
			{Gingras}},\ and\ \bibinfo {author} {\bibfnamefont {J.~E.}\ \bibnamefont
			{Greedan}},\ }\bibfield  {title} {\bibinfo {title} {Magnetic pyrochlore
			oxides},\ }\href {https://doi.org/10.1103/RevModPhys.82.53} {\bibfield
		{journal} {\bibinfo  {journal} {Rev. Mod. Phys.}\ }\textbf {\bibinfo {volume}
			{82}},\ \bibinfo {pages} {53} (\bibinfo {year} {2010})}\BibitemShut {NoStop}%
	\bibitem [{\citenamefont {Jin}\ and\ \citenamefont {Zhou}(2020)}]{Jin054408}%
	\BibitemOpen
	\bibfield  {author} {\bibinfo {author} {\bibfnamefont {H.-K.}\ \bibnamefont
			{Jin}}\ and\ \bibinfo {author} {\bibfnamefont {Y.}~\bibnamefont {Zhou}},\
	}\bibfield  {title} {\bibinfo {title} {Classical and quantum order in
			hyperkagome antiferromagnets},\ }\href
	{https://doi.org/10.1103/PhysRevB.101.054408} {\bibfield  {journal} {\bibinfo
			{journal} {Phys. Rev. B}\ }\textbf {\bibinfo {volume} {101}},\ \bibinfo
		{pages} {054408} (\bibinfo {year} {2020})}\BibitemShut {NoStop}%
	\bibitem [{\citenamefont {Geller}(1967)}]{Geller1}%
	\BibitemOpen
	\bibfield  {author} {\bibinfo {author} {\bibfnamefont {S.}~\bibnamefont
			{Geller}},\ }\bibfield  {title} {\bibinfo {title} {{Crystal chemistry of the
				garnets}},\ }\href {https://doi.org/doi:10.1524/zkri.1967.125.16.1}
	{\bibfield  {journal} {\bibinfo  {journal} {Zeitschrift für
				Kristallographie}\ }\textbf {\bibinfo {volume} {125}},\ \bibinfo {pages} {1}
		(\bibinfo {year} {1967})}\BibitemShut {NoStop}%
	\bibitem [{\citenamefont {Cui}\ \emph {et~al.}(2020)\citenamefont {Cui},
		\citenamefont {Huang}, \citenamefont {Alonso}, \citenamefont {Sheptyakov},
		\citenamefont {De~la Cruz}, \citenamefont {Fern\'andez-D\'{\i}az},
		\citenamefont {Wang}, \citenamefont {Cai}, \citenamefont {Li}, \citenamefont
		{Dong}, \citenamefont {Zhou},\ and\ \citenamefont {Cheng}}]{Cui144424}%
	\BibitemOpen
	\bibfield  {author} {\bibinfo {author} {\bibfnamefont {Q.}~\bibnamefont
			{Cui}}, \bibinfo {author} {\bibfnamefont {Q.}~\bibnamefont {Huang}}, \bibinfo
		{author} {\bibfnamefont {J.~A.}\ \bibnamefont {Alonso}}, \bibinfo {author}
		{\bibfnamefont {D.}~\bibnamefont {Sheptyakov}}, \bibinfo {author}
		{\bibfnamefont {C.~R.}\ \bibnamefont {De~la Cruz}}, \bibinfo {author}
		{\bibfnamefont {M.~T.}\ \bibnamefont {Fern\'andez-D\'{\i}az}}, \bibinfo
		{author} {\bibfnamefont {N.~N.}\ \bibnamefont {Wang}}, \bibinfo {author}
		{\bibfnamefont {Y.~Q.}\ \bibnamefont {Cai}}, \bibinfo {author} {\bibfnamefont
			{D.}~\bibnamefont {Li}}, \bibinfo {author} {\bibfnamefont {X.~L.}\
			\bibnamefont {Dong}}, \bibinfo {author} {\bibfnamefont {H.~D.}\ \bibnamefont
			{Zhou}},\ and\ \bibinfo {author} {\bibfnamefont {J.-G.}\ \bibnamefont
			{Cheng}},\ }\bibfield  {title} {\bibinfo {title} {{Complex antiferromagnetic
				order in the garnet
				$\mathrm{C}{\mathrm{o}}_{3}\mathrm{A}{\mathrm{l}}_{2}\mathrm{S}{\mathrm{i}}_{3}{\mathrm{O}}_{12}$}},\
	}\href {https://doi.org/10.1103/PhysRevB.101.144424} {\bibfield  {journal}
		{\bibinfo  {journal} {Phys. Rev. B}\ }\textbf {\bibinfo {volume} {101}},\
		\bibinfo {pages} {144424} (\bibinfo {year} {2020})}\BibitemShut {NoStop}%
	\bibitem [{\citenamefont {Min}\ \emph {et~al.}(2022)\citenamefont {Min},
		\citenamefont {Zheng}, \citenamefont {Gong}, \citenamefont {Chen},
		\citenamefont {Liu}, \citenamefont {Xie}, \citenamefont {Zhang},
		\citenamefont {Ma}, \citenamefont {Liu}, \citenamefont {Wang}, \citenamefont
		{Li},\ and\ \citenamefont {Liu}}]{Min86}%
	\BibitemOpen
	\bibfield  {author} {\bibinfo {author} {\bibfnamefont {J.}~\bibnamefont
			{Min}}, \bibinfo {author} {\bibfnamefont {S.}~\bibnamefont {Zheng}}, \bibinfo
		{author} {\bibfnamefont {J.}~\bibnamefont {Gong}}, \bibinfo {author}
		{\bibfnamefont {X.}~\bibnamefont {Chen}}, \bibinfo {author} {\bibfnamefont
			{F.}~\bibnamefont {Liu}}, \bibinfo {author} {\bibfnamefont {Y.}~\bibnamefont
			{Xie}}, \bibinfo {author} {\bibfnamefont {Y.}~\bibnamefont {Zhang}}, \bibinfo
		{author} {\bibfnamefont {Z.}~\bibnamefont {Ma}}, \bibinfo {author}
		{\bibfnamefont {M.}~\bibnamefont {Liu}}, \bibinfo {author} {\bibfnamefont
			{X.}~\bibnamefont {Wang}}, \bibinfo {author} {\bibfnamefont {H.}~\bibnamefont
			{Li}},\ and\ \bibinfo {author} {\bibfnamefont {J.-M.}\ \bibnamefont {Liu}},\
	}\bibfield  {title} {\bibinfo {title} {{Magnetoelectric Effect in Garnet
				${\mathrm{Mn}}_{3}{\mathrm{Al}}_{2}{\mathrm{Ge}}_{3}{\mathrm{O}}_{12}$}},\
	}\href {https://doi.org/10.1021/acs.inorgchem.1c01935} {\bibfield  {journal}
		{\bibinfo  {journal} {Inorg. Chem.}\ }\textbf {\bibinfo {volume} {61}},\
		\bibinfo {pages} {86} (\bibinfo {year} {2022})}\BibitemShut {NoStop}%
	\bibitem [{\citenamefont {Paddison}\ \emph {et~al.}(2015)\citenamefont
		{Paddison}, \citenamefont {Jacobsen}, \citenamefont {Petrenko}, \citenamefont
		{Fernandez-Diaz}, \citenamefont {Deen},\ and\ \citenamefont
		{Goodwin}}]{Paddison179}%
	\BibitemOpen
	\bibfield  {author} {\bibinfo {author} {\bibfnamefont {J.~A.~M.}\
			\bibnamefont {Paddison}}, \bibinfo {author} {\bibfnamefont {H.}~\bibnamefont
			{Jacobsen}}, \bibinfo {author} {\bibfnamefont {O.~A.}\ \bibnamefont
			{Petrenko}}, \bibinfo {author} {\bibfnamefont {M.~T.}\ \bibnamefont
			{Fernandez-Diaz}}, \bibinfo {author} {\bibfnamefont {P.~P.}\ \bibnamefont
			{Deen}},\ and\ \bibinfo {author} {\bibfnamefont {A.~L.}\ \bibnamefont
			{Goodwin}},\ }\bibfield  {title} {\bibinfo {title} {{Hidden order in
				spin-liquid Gd$_3$Ga$_5$O$_{12}$}},\ }\href
	{https://doi.org/10.1126/science.aaa5326} {\bibfield  {journal} {\bibinfo
			{journal} {Science}\ }\textbf {\bibinfo {volume} {350}},\ \bibinfo {pages}
		{179} (\bibinfo {year} {2015})}\BibitemShut {NoStop}%
	\bibitem [{\citenamefont {Kamazawa}\ \emph {et~al.}(2008)\citenamefont
		{Kamazawa}, \citenamefont {Louca}, \citenamefont {Morinaga}, \citenamefont
		{Sato}, \citenamefont {Huang}, \citenamefont {Copley},\ and\ \citenamefont
		{Qiu}}]{Kamazawa064412}%
	\BibitemOpen
	\bibfield  {author} {\bibinfo {author} {\bibfnamefont {K.}~\bibnamefont
			{Kamazawa}}, \bibinfo {author} {\bibfnamefont {D.}~\bibnamefont {Louca}},
		\bibinfo {author} {\bibfnamefont {R.}~\bibnamefont {Morinaga}}, \bibinfo
		{author} {\bibfnamefont {T.~J.}\ \bibnamefont {Sato}}, \bibinfo {author}
		{\bibfnamefont {Q.}~\bibnamefont {Huang}}, \bibinfo {author} {\bibfnamefont
			{J.~R.~D.}\ \bibnamefont {Copley}},\ and\ \bibinfo {author} {\bibfnamefont
			{Y.}~\bibnamefont {Qiu}},\ }\bibfield  {title} {\bibinfo {title}
		{{Field-induced antiferromagnetism and competition in the metamagnetic state
				of terbium gallium garnet}},\ }\href
	{https://doi.org/10.1103/PhysRevB.78.064412} {\bibfield  {journal} {\bibinfo
			{journal} {Phys. Rev. B}\ }\textbf {\bibinfo {volume} {78}},\ \bibinfo
		{pages} {064412} (\bibinfo {year} {2008})}\BibitemShut {NoStop}%
	\bibitem [{\citenamefont {Wawrzy\ifmmode~\acute{n}\else \'{n}\fi{}czak}\ \emph
		{et~al.}(2019)\citenamefont {Wawrzy\ifmmode~\acute{n}\else \'{n}\fi{}czak},
		\citenamefont {Tomasello}, \citenamefont {Manuel}, \citenamefont {Khalyavin},
		\citenamefont {Le}, \citenamefont {Guidi}, \citenamefont {Cervellino},
		\citenamefont {Ziman}, \citenamefont {Boehm}, \citenamefont {Nilsen},\ and\
		\citenamefont {Fennell}}]{Wawrzy094442}%
	\BibitemOpen
	\bibfield  {author} {\bibinfo {author} {\bibfnamefont {R.}~\bibnamefont
			{Wawrzy\ifmmode~\acute{n}\else \'{n}\fi{}czak}}, \bibinfo {author}
		{\bibfnamefont {B.}~\bibnamefont {Tomasello}}, \bibinfo {author}
		{\bibfnamefont {P.}~\bibnamefont {Manuel}}, \bibinfo {author} {\bibfnamefont
			{D.}~\bibnamefont {Khalyavin}}, \bibinfo {author} {\bibfnamefont {M.~D.}\
			\bibnamefont {Le}}, \bibinfo {author} {\bibfnamefont {T.}~\bibnamefont
			{Guidi}}, \bibinfo {author} {\bibfnamefont {A.}~\bibnamefont {Cervellino}},
		\bibinfo {author} {\bibfnamefont {T.}~\bibnamefont {Ziman}}, \bibinfo
		{author} {\bibfnamefont {M.}~\bibnamefont {Boehm}}, \bibinfo {author}
		{\bibfnamefont {G.~J.}\ \bibnamefont {Nilsen}},\ and\ \bibinfo {author}
		{\bibfnamefont {T.}~\bibnamefont {Fennell}},\ }\bibfield  {title} {\bibinfo
		{title} {{Magnetic order and single-ion anisotropy in
				${\mathrm{Tb}}_{3}{\mathrm{Ga}}_{5}{\mathrm{O}}_{12}$}},\ }\href
	{https://doi.org/10.1103/PhysRevB.100.094442} {\bibfield  {journal} {\bibinfo
			{journal} {Phys. Rev. B}\ }\textbf {\bibinfo {volume} {100}},\ \bibinfo
		{pages} {094442} (\bibinfo {year} {2019})}\BibitemShut {NoStop}%
	\bibitem [{\citenamefont {Zhou}\ \emph {et~al.}(2008)\citenamefont {Zhou},
		\citenamefont {Wiebe}, \citenamefont {Balicas}, \citenamefont {Yo},
		\citenamefont {Qiu}, \citenamefont {Copley},\ and\ \citenamefont
		{Gardner}}]{Zhou140406}%
	\BibitemOpen
	\bibfield  {author} {\bibinfo {author} {\bibfnamefont {H.~D.}\ \bibnamefont
			{Zhou}}, \bibinfo {author} {\bibfnamefont {C.~R.}\ \bibnamefont {Wiebe}},
		\bibinfo {author} {\bibfnamefont {L.}~\bibnamefont {Balicas}}, \bibinfo
		{author} {\bibfnamefont {Y.~J.}\ \bibnamefont {Yo}}, \bibinfo {author}
		{\bibfnamefont {Y.}~\bibnamefont {Qiu}}, \bibinfo {author} {\bibfnamefont
			{J.~R.~D.}\ \bibnamefont {Copley}},\ and\ \bibinfo {author} {\bibfnamefont
			{J.~S.}\ \bibnamefont {Gardner}},\ }\bibfield  {title} {\bibinfo {title}
		{{Intrinsic spin-disordered ground state of the Ising garnet
				${\text{Ho}}_{3}{\text{Ga}}_{5}{\text{O}}_{12}$}},\ }\href
	{https://doi.org/10.1103/PhysRevB.78.140406} {\bibfield  {journal} {\bibinfo
			{journal} {Phys. Rev. B}\ }\textbf {\bibinfo {volume} {78}},\ \bibinfo
		{pages} {140406} (\bibinfo {year} {2008})}\BibitemShut {NoStop}%
	\bibitem [{\citenamefont {Belov}\ and\ \citenamefont
		{Sokolov}(1977)}]{Konstantin149}%
	\BibitemOpen
	\bibfield  {author} {\bibinfo {author} {\bibfnamefont {K.~P.}\ \bibnamefont
			{Belov}}\ and\ \bibinfo {author} {\bibfnamefont {V.~I.}\ \bibnamefont
			{Sokolov}},\ }\bibfield  {title} {\bibinfo {title} {{Antiferromagnetic
				garnets}},\ }\href {https://doi.org/10.1070/PU1977v020n02ABEH005332}
	{\bibfield  {journal} {\bibinfo  {journal} {Sov. Phys. Usp.}\ }\textbf
		{\bibinfo {volume} {20}},\ \bibinfo {pages} {149} (\bibinfo {year}
		{1977})}\BibitemShut {NoStop}%
	\bibitem [{\citenamefont {Kohara}\ \emph {et~al.}(2010)\citenamefont {Kohara},
		\citenamefont {Yamasaki}, \citenamefont {Onose},\ and\ \citenamefont
		{Tokura}}]{Kohara104419}%
	\BibitemOpen
	\bibfield  {author} {\bibinfo {author} {\bibfnamefont {Y.}~\bibnamefont
			{Kohara}}, \bibinfo {author} {\bibfnamefont {Y.}~\bibnamefont {Yamasaki}},
		\bibinfo {author} {\bibfnamefont {Y.}~\bibnamefont {Onose}},\ and\ \bibinfo
		{author} {\bibfnamefont {Y.}~\bibnamefont {Tokura}},\ }\bibfield  {title}
	{\bibinfo {title} {{Excess-electron induced polarization and magnetoelectric
				effect in yttrium iron garnet}},\ }\href
	{https://doi.org/10.1103/PhysRevB.82.104419} {\bibfield  {journal} {\bibinfo
			{journal} {Phys. Rev. B}\ }\textbf {\bibinfo {volume} {82}},\ \bibinfo
		{pages} {104419} (\bibinfo {year} {2010})}\BibitemShut {NoStop}%
	\bibitem [{\citenamefont {Barker}\ and\ \citenamefont
		{Bauer}(2016)}]{Barker217201}%
	\BibitemOpen
	\bibfield  {author} {\bibinfo {author} {\bibfnamefont {J.}~\bibnamefont
			{Barker}}\ and\ \bibinfo {author} {\bibfnamefont {G.~E.~W.}\ \bibnamefont
			{Bauer}},\ }\bibfield  {title} {\bibinfo {title} {{Thermal Spin Dynamics of
				Yttrium Iron Garnet}},\ }\href
	{https://doi.org/10.1103/PhysRevLett.117.217201} {\bibfield  {journal}
		{\bibinfo  {journal} {Phys. Rev. Lett.}\ }\textbf {\bibinfo {volume} {117}},\
		\bibinfo {pages} {217201} (\bibinfo {year} {2016})}\BibitemShut {NoStop}%
	\bibitem [{\citenamefont {Bozorth}\ and\ \citenamefont
		{Geller}(1959)}]{Bozorth263}%
	\BibitemOpen
	\bibfield  {author} {\bibinfo {author} {\bibfnamefont {R.}~\bibnamefont
			{Bozorth}}\ and\ \bibinfo {author} {\bibfnamefont {S.}~\bibnamefont
			{Geller}},\ }\bibfield  {title} {\bibinfo {title} {{Interactions and
				distributions of magnetic ions in some garnet systems}},\ }\href
	{https://doi.org/https://doi.org/10.1016/0022-3697(59)90224-0} {\bibfield
		{journal} {\bibinfo  {journal} {J. Phys. Chem. Solids}\ }\textbf {\bibinfo
			{volume} {11}},\ \bibinfo {pages} {263} (\bibinfo {year} {1959})}\BibitemShut
	{NoStop}%
	\bibitem [{\citenamefont {Belov}\ \emph {et~al.}(1972)\citenamefont {Belov},
		\citenamefont {Mamsurova}, \citenamefont {Mill},\ and\ \citenamefont
		{Sokolov}}]{Belov173}%
	\BibitemOpen
	\bibfield  {author} {\bibinfo {author} {\bibfnamefont {K.~P.}\ \bibnamefont
			{Belov}}, \bibinfo {author} {\bibfnamefont {D.~G.}\ \bibnamefont
			{Mamsurova}}, \bibinfo {author} {\bibfnamefont {B.~V.}\ \bibnamefont
			{Mill}},\ and\ \bibinfo {author} {\bibfnamefont {V.~I.}\ \bibnamefont
			{Sokolov}},\ }\bibfield  {title} {\bibinfo {title} {{Ferromagnetism of the
				garnet
				${\mathrm{Mn}}_{3}{\mathrm{Cr}}_{2}{\mathrm{Ge}}_{3}{\mathrm{O}}_{12}$}},\
	}\href {http://jetpletters.ru/ps/1759/article_26762.pdf} {\bibfield
		{journal} {\bibinfo  {journal} {JETP Lett.}\ }\textbf {\bibinfo {volume}
			{16}},\ \bibinfo {pages} {120} (\bibinfo {year} {1972})}\BibitemShut
	{NoStop}%
	\bibitem [{\citenamefont {Valyanskaya}\ and\ \citenamefont
		{Sokolov}(1978)}]{Valyanskaya2114}%
	\BibitemOpen
	\bibfield  {author} {\bibinfo {author} {\bibfnamefont {T.~V.}\ \bibnamefont
			{Valyanskaya}}\ and\ \bibinfo {author} {\bibfnamefont {V.~I.}\ \bibnamefont
			{Sokolov}},\ }\bibfield  {title} {\bibinfo {title} {{Features of
				antiferromagnetic ordering in the garnet
				${\mathrm{Mn}}_{3}{\mathrm{Cr}}_{2}{\mathrm{Ge}}_{3}{\mathrm{O}}_{12}$}},\
	}\href {https://doi.org/http://www.jetp.ras.ru/cgi-bin/dn/e_048_01_0161.pdf}
	{\bibfield  {journal} {\bibinfo  {journal} {Sov. Phys. JETP}\ }\textbf
		{\bibinfo {volume} {75}},\ \bibinfo {pages} {161} (\bibinfo {year}
		{1978})}\BibitemShut {NoStop}%
	\bibitem [{\citenamefont {Tokiwa}\ \emph {et~al.}(2021)\citenamefont {Tokiwa},
		\citenamefont {Bachus}, \citenamefont {Kavita}, \citenamefont {Jesche},
		\citenamefont {Tsirlin},\ and\ \citenamefont {Gegenwart}}]{Tokiwa42}%
	\BibitemOpen
	\bibfield  {author} {\bibinfo {author} {\bibfnamefont {Y.}~\bibnamefont
			{Tokiwa}}, \bibinfo {author} {\bibfnamefont {S.}~\bibnamefont {Bachus}},
		\bibinfo {author} {\bibfnamefont {K.}~\bibnamefont {Kavita}}, \bibinfo
		{author} {\bibfnamefont {A.}~\bibnamefont {Jesche}}, \bibinfo {author}
		{\bibfnamefont {A.~A.}\ \bibnamefont {Tsirlin}},\ and\ \bibinfo {author}
		{\bibfnamefont {P.}~\bibnamefont {Gegenwart}},\ }\bibfield  {title} {\bibinfo
		{title} {{Frustrated magnet for adiabatic demagnetization cooling to
				milli-Kelvin temperatures}},\ }\href
	{https://doi.org/10.1038/s43246-021-00142-1} {\bibfield  {journal} {\bibinfo
			{journal} {Commun. Mater.}\ }\textbf {\bibinfo {volume} {2}},\ \bibinfo
		{pages} {42} (\bibinfo {year} {2021})}\BibitemShut {NoStop}%
	\bibitem [{\citenamefont {Zhitomirsky}(2003)}]{Zhitomirsky104421}%
	\BibitemOpen
	\bibfield  {author} {\bibinfo {author} {\bibfnamefont {M.~E.}\ \bibnamefont
			{Zhitomirsky}},\ }\bibfield  {title} {\bibinfo {title} {{Enhanced
				magnetocaloric effect in frustrated magnets}},\ }\href
	{https://doi.org/10.1103/PhysRevB.67.104421} {\bibfield  {journal} {\bibinfo
			{journal} {Phys. Rev. B}\ }\textbf {\bibinfo {volume} {67}},\ \bibinfo
		{pages} {104421} (\bibinfo {year} {2003})}\BibitemShut {NoStop}%
	\bibitem [{\citenamefont {Kleinhans}\ \emph {et~al.}(2023)\citenamefont
		{Kleinhans}, \citenamefont {Eibensteiner}, \citenamefont {Leiner},
		\citenamefont {Resch}, \citenamefont {Worch}, \citenamefont {Wilde},
		\citenamefont {Spallek}, \citenamefont {Regnat},\ and\ \citenamefont
		{Pfleiderer}}]{Kleinhans014038}%
	\BibitemOpen
	\bibfield  {author} {\bibinfo {author} {\bibfnamefont {M.}~\bibnamefont
			{Kleinhans}}, \bibinfo {author} {\bibfnamefont {K.}~\bibnamefont
			{Eibensteiner}}, \bibinfo {author} {\bibfnamefont {J.}~\bibnamefont
			{Leiner}}, \bibinfo {author} {\bibfnamefont {C.}~\bibnamefont {Resch}},
		\bibinfo {author} {\bibfnamefont {L.}~\bibnamefont {Worch}}, \bibinfo
		{author} {\bibfnamefont {M.}~\bibnamefont {Wilde}}, \bibinfo {author}
		{\bibfnamefont {J.}~\bibnamefont {Spallek}}, \bibinfo {author} {\bibfnamefont
			{A.}~\bibnamefont {Regnat}},\ and\ \bibinfo {author} {\bibfnamefont
			{C.}~\bibnamefont {Pfleiderer}},\ }\bibfield  {title} {\bibinfo {title}
		{{Magnetocaloric Properties of ${R}_{3}{\mathrm{Ga}}_{5}{\mathrm{O}}_{12}$
				($R=\text{Tb, Gd, Nd, Dy}$)}},\ }\href
	{https://doi.org/10.1103/PhysRevApplied.19.014038} {\bibfield  {journal}
		{\bibinfo  {journal} {Phys. Rev. Appl.}\ }\textbf {\bibinfo {volume} {19}},\
		\bibinfo {pages} {014038} (\bibinfo {year} {2023})}\BibitemShut {NoStop}%
	\bibitem [{\citenamefont {Urata}\ \emph {et~al.}(2001)\citenamefont {Urata},
		\citenamefont {Wada}, \citenamefont {Tashiro},\ and\ \citenamefont
		{Deng}}]{Urata801}%
	\BibitemOpen
	\bibfield  {author} {\bibinfo {author} {\bibfnamefont {Y.}~\bibnamefont
			{Urata}}, \bibinfo {author} {\bibfnamefont {S.}~\bibnamefont {Wada}},
		\bibinfo {author} {\bibfnamefont {H.}~\bibnamefont {Tashiro}},\ and\ \bibinfo
		{author} {\bibfnamefont {P.}~\bibnamefont {Deng}},\ }\bibfield  {title}
	{\bibinfo {title} {{Laser performance of highly neodymium-doped yttrium
				aluminum garnet crystals}},\ }\href {https://doi.org/10.1364/OL.26.000801}
	{\bibfield  {journal} {\bibinfo  {journal} {Opt. Lett.}\ }\textbf {\bibinfo
			{volume} {26}},\ \bibinfo {pages} {801} (\bibinfo {year} {2001})}\BibitemShut
	{NoStop}%
	\bibitem [{\citenamefont {Lipp}\ \emph {et~al.}(2012)\citenamefont {Lipp},
		\citenamefont {Strobel}, \citenamefont {Lissner},\ and\ \citenamefont
		{Niewa}}]{Lippi35}%
	\BibitemOpen
	\bibfield  {author} {\bibinfo {author} {\bibfnamefont {C.}~\bibnamefont
			{Lipp}}, \bibinfo {author} {\bibfnamefont {S.}~\bibnamefont {Strobel}},
		\bibinfo {author} {\bibfnamefont {F.}~\bibnamefont {Lissner}},\ and\ \bibinfo
		{author} {\bibfnamefont {R.}~\bibnamefont {Niewa}},\ }\bibfield  {title}
	{\bibinfo {title} {{Garnet-type Mn${\sb 3}$Cr${\sb 2}$(GeO${\sb 4}$)${\sb
					3}$}},\ }\href {https://doi.org/10.1107/S1600536812016832} {\bibfield
		{journal} {\bibinfo  {journal} {Acta Cryst. E}\ }\textbf {\bibinfo {volume}
			{68}},\ \bibinfo {pages} {i35} (\bibinfo {year} {2012})}\BibitemShut
	{NoStop}%
	\bibitem [{\citenamefont {Carvajal}(1993)}]{Carvajal55}%
	\BibitemOpen
	\bibfield  {author} {\bibinfo {author} {\bibfnamefont {J.~R.}\ \bibnamefont
			{Carvajal}},\ }\bibfield  {title} {\bibinfo {title} {Recent advances in
			magnetic structure determination by neutron powder diffraction},\ }\href
	{https://doi.org/https://doi.org/10.1016/0921-4526(93)90108-I} {\bibfield
		{journal} {\bibinfo  {journal} {Physica B: Condens. Matter}\ }\textbf
		{\bibinfo {volume} {192}},\ \bibinfo {pages} {55} (\bibinfo {year}
		{1993})}\BibitemShut {NoStop}%
	\bibitem [{\citenamefont {Xiang}\ \emph {et~al.}(2011)\citenamefont {Xiang},
		\citenamefont {Kan}, \citenamefont {Wei}, \citenamefont {Whangbo},\ and\
		\citenamefont {Gong}}]{xiang2011}%
	\BibitemOpen
	\bibfield  {author} {\bibinfo {author} {\bibfnamefont {H.~J.}\ \bibnamefont
			{Xiang}}, \bibinfo {author} {\bibfnamefont {E.~J.}\ \bibnamefont {Kan}},
		\bibinfo {author} {\bibfnamefont {S.-H.}\ \bibnamefont {Wei}}, \bibinfo
		{author} {\bibfnamefont {M.-H.}\ \bibnamefont {Whangbo}},\ and\ \bibinfo
		{author} {\bibfnamefont {X.~G.}\ \bibnamefont {Gong}},\ }\bibfield  {title}
	{\bibinfo {title} {Predicting the spin-lattice order of frustrated systems
			from first principles},\ }\href {https://doi.org/10.1103/PhysRevB.84.224429}
	{\bibfield  {journal} {\bibinfo  {journal} {Phys. Rev. B}\ }\textbf {\bibinfo
			{volume} {84}},\ \bibinfo {pages} {224429} (\bibinfo {year}
		{2011})}\BibitemShut {NoStop}%
	\bibitem [{\citenamefont {Kresse}\ and\ \citenamefont
		{Furthm\"uller}(1996{\natexlab{a}})}]{vasp1}%
	\BibitemOpen
	\bibfield  {author} {\bibinfo {author} {\bibfnamefont {G.}~\bibnamefont
			{Kresse}}\ and\ \bibinfo {author} {\bibfnamefont {J.}~\bibnamefont
			{Furthm\"uller}},\ }\bibfield  {title} {\bibinfo {title} {Efficiency of
			ab-initio total energy calculations for metals and semiconductors using a
			plane-wave basis set},\ }\href {https://doi.org/10.1016/0927-0256(96)00008-0}
	{\bibfield  {journal} {\bibinfo  {journal} {Comput. Mater. Sci.}\ }\textbf
		{\bibinfo {volume} {6}},\ \bibinfo {pages} {15} (\bibinfo {year}
		{1996}{\natexlab{a}})}\BibitemShut {NoStop}%
	\bibitem [{\citenamefont {Kresse}\ and\ \citenamefont
		{Furthm\"uller}(1996{\natexlab{b}})}]{vasp2}%
	\BibitemOpen
	\bibfield  {author} {\bibinfo {author} {\bibfnamefont {G.}~\bibnamefont
			{Kresse}}\ and\ \bibinfo {author} {\bibfnamefont {J.}~\bibnamefont
			{Furthm\"uller}},\ }\bibfield  {title} {\bibinfo {title} {Efficient iterative
			schemes for \textit{ab initio} total-energy calculations using a plane-wave
			basis set},\ }\href {https://doi.org/10.1103/PhysRevB.54.11169} {\bibfield
		{journal} {\bibinfo  {journal} {Phys. Rev. B}\ }\textbf {\bibinfo {volume}
			{54}},\ \bibinfo {pages} {11169} (\bibinfo {year}
		{1996}{\natexlab{b}})}\BibitemShut {NoStop}%
	\bibitem [{\citenamefont {Perdew}\ \emph {et~al.}(1996)\citenamefont {Perdew},
		\citenamefont {Burke},\ and\ \citenamefont {Ernzerhof}}]{pbe96}%
	\BibitemOpen
	\bibfield  {author} {\bibinfo {author} {\bibfnamefont {J.~P.}\ \bibnamefont
			{Perdew}}, \bibinfo {author} {\bibfnamefont {K.}~\bibnamefont {Burke}},\ and\
		\bibinfo {author} {\bibfnamefont {M.}~\bibnamefont {Ernzerhof}},\ }\bibfield
	{title} {\bibinfo {title} {Generalized gradient approximation made simple},\
	}\href {https://doi.org/10.1103/PhysRevLett.77.3865} {\bibfield  {journal}
		{\bibinfo  {journal} {Phys. Rev. Lett.}\ }\textbf {\bibinfo {volume} {77}},\
		\bibinfo {pages} {3865} (\bibinfo {year} {1996})}\BibitemShut {NoStop}%
	\bibitem [{\citenamefont {Nath}\ \emph {et~al.}(2014)\citenamefont {Nath},
		\citenamefont {Ranjith}, \citenamefont {Roy}, \citenamefont {Johnston},
		\citenamefont {Furukawa},\ and\ \citenamefont {Tsirlin}}]{nath2014}%
	\BibitemOpen
	\bibfield  {author} {\bibinfo {author} {\bibfnamefont {R.}~\bibnamefont
			{Nath}}, \bibinfo {author} {\bibfnamefont {K.~M.}\ \bibnamefont {Ranjith}},
		\bibinfo {author} {\bibfnamefont {B.}~\bibnamefont {Roy}}, \bibinfo {author}
		{\bibfnamefont {D.~C.}\ \bibnamefont {Johnston}}, \bibinfo {author}
		{\bibfnamefont {Y.}~\bibnamefont {Furukawa}},\ and\ \bibinfo {author}
		{\bibfnamefont {A.~A.}\ \bibnamefont {Tsirlin}},\ }\bibfield  {title}
	{\bibinfo {title} {Magnetic transitions in the spin-5/2 frustrated magnet
			{BiMn$_2$PO$_6$} and strong lattice softening in {BiMn$_2$PO$_6$} and
			{BiZn$_2$PO$_6$ below 200 K}},\ }\href
	{https://doi.org/10.1103/PhysRevB.90.024431} {\bibfield  {journal} {\bibinfo
			{journal} {Phys. Rev. B}\ }\textbf {\bibinfo {volume} {90}},\ \bibinfo
		{pages} {024431} (\bibinfo {year} {2014})}\BibitemShut {NoStop}%
	\bibitem [{\citenamefont {Janson}\ \emph {et~al.}(2014)\citenamefont {Janson},
		\citenamefont {N\'enert}, \citenamefont {Isobe}, \citenamefont {Skourski},
		\citenamefont {Ueda}, \citenamefont {Rosner},\ and\ \citenamefont
		{Tsirlin}}]{janson2014}%
	\BibitemOpen
	\bibfield  {author} {\bibinfo {author} {\bibfnamefont {O.}~\bibnamefont
			{Janson}}, \bibinfo {author} {\bibfnamefont {G.}~\bibnamefont {N\'enert}},
		\bibinfo {author} {\bibfnamefont {M.}~\bibnamefont {Isobe}}, \bibinfo
		{author} {\bibfnamefont {Y.}~\bibnamefont {Skourski}}, \bibinfo {author}
		{\bibfnamefont {Y.}~\bibnamefont {Ueda}}, \bibinfo {author} {\bibfnamefont
			{H.}~\bibnamefont {Rosner}},\ and\ \bibinfo {author} {\bibfnamefont {A.~A.}\
			\bibnamefont {Tsirlin}},\ }\bibfield  {title} {\bibinfo {title} {Magnetic
			pyroxenes {LiCrGe$_2$O$_6$ and LiCrSi$_2$O$_6$}: Dimensionality crossover in
			a nonfrustrated {$S=\frac32$} {Heisenberg} model},\ }\href
	{https://doi.org/10.1103/PhysRevB.90.214424} {\bibfield  {journal} {\bibinfo
			{journal} {Phys. Rev. B}\ }\textbf {\bibinfo {volume} {90}},\ \bibinfo
		{pages} {214424} (\bibinfo {year} {2014})}\BibitemShut {NoStop}%
	\bibitem [{\citenamefont {Mohanty}\ \emph {et~al.}(2023)\citenamefont
		{Mohanty}, \citenamefont {Babu}, \citenamefont {Furukawa},\ and\
		\citenamefont {Nath}}]{Mohanty104424}%
	\BibitemOpen
	\bibfield  {author} {\bibinfo {author} {\bibfnamefont {S.}~\bibnamefont
			{Mohanty}}, \bibinfo {author} {\bibfnamefont {J.}~\bibnamefont {Babu}},
		\bibinfo {author} {\bibfnamefont {Y.}~\bibnamefont {Furukawa}},\ and\
		\bibinfo {author} {\bibfnamefont {R.}~\bibnamefont {Nath}},\ }\bibfield
	{title} {\bibinfo {title} {{Structural and double magnetic transitions in the
				frustrated spin-$\frac{1}{2}$ capped-kagome antiferromagnet
				$(\mathrm{RbCl}){\mathrm{Cu}}_{5}{\mathrm{P}}_{2}{\mathrm{O}}_{10}$}},\
	}\href {https://doi.org/10.1103/PhysRevB.108.104424} {\bibfield  {journal}
		{\bibinfo  {journal} {Phys. Rev. B}\ }\textbf {\bibinfo {volume} {108}},\
		\bibinfo {pages} {104424} (\bibinfo {year} {2023})}\BibitemShut {NoStop}%
	\bibitem [{\citenamefont {Nath}\ \emph {et~al.}(2010)\citenamefont {Nath},
		\citenamefont {Garlea}, \citenamefont {Goldman},\ and\ \citenamefont
		{Johnston}}]{Nath224513}%
	\BibitemOpen
	\bibfield  {author} {\bibinfo {author} {\bibfnamefont {R.}~\bibnamefont
			{Nath}}, \bibinfo {author} {\bibfnamefont {V.~O.}\ \bibnamefont {Garlea}},
		\bibinfo {author} {\bibfnamefont {A.~I.}\ \bibnamefont {Goldman}},\ and\
		\bibinfo {author} {\bibfnamefont {D.~C.}\ \bibnamefont {Johnston}},\
	}\bibfield  {title} {\bibinfo {title} {{Synthesis, structure, and properties
				of tetragonal ${\text{Sr}}_{2}{M}_{3}{\text{As}}_{2}{\text{O}}_{2}$
				(${M}_{3}={\text{Mn}}_{3}$, ${\text{Mn}}_{2}\text{Cu}$, and
				${\text{MnZn}}_{2}$) compounds containing alternating ${\text{CuO}}_{2}$-type
				and FeAs-type layers}},\ }\href {https://doi.org/10.1103/PhysRevB.81.224513}
	{\bibfield  {journal} {\bibinfo  {journal} {Phys. Rev. B}\ }\textbf {\bibinfo
			{volume} {81}},\ \bibinfo {pages} {224513} (\bibinfo {year}
		{2010})}\BibitemShut {NoStop}%
	\bibitem [{\citenamefont {Subramanian}\ \emph {et~al.}(1988)\citenamefont
		{Subramanian}, \citenamefont {Torardi}, \citenamefont {Johnson},
		\citenamefont {Pannetier},\ and\ \citenamefont {Sleight}}]{Subramanian24}%
	\BibitemOpen
	\bibfield  {author} {\bibinfo {author} {\bibfnamefont {M.}~\bibnamefont
			{Subramanian}}, \bibinfo {author} {\bibfnamefont {C.}~\bibnamefont
			{Torardi}}, \bibinfo {author} {\bibfnamefont {D.}~\bibnamefont {Johnson}},
		\bibinfo {author} {\bibfnamefont {J.}~\bibnamefont {Pannetier}},\ and\
		\bibinfo {author} {\bibfnamefont {A.}~\bibnamefont {Sleight}},\ }\bibfield
	{title} {\bibinfo {title} {{Ferromagnetic $R_2$Mn$_2$O$_7$ pyrochlores ($R$ =
				Dy, Lu, Y)}},\ }\href
	{https://doi.org/https://doi.org/10.1016/0022-4596(88)90004-7} {\bibfield
		{journal} {\bibinfo  {journal} {J. Solid State Chem.}\ }\textbf {\bibinfo
			{volume} {72}},\ \bibinfo {pages} {24} (\bibinfo {year} {1988})}\BibitemShut
	{NoStop}%
	\bibitem [{\citenamefont {Savina}\ \emph {et~al.}(2011)\citenamefont {Savina},
		\citenamefont {Bludov}, \citenamefont {Pashchenko}, \citenamefont
		{Gnatchenko}, \citenamefont {Lemmens},\ and\ \citenamefont
		{Berger}}]{Savina104447}%
	\BibitemOpen
	\bibfield  {author} {\bibinfo {author} {\bibfnamefont {Y.}~\bibnamefont
			{Savina}}, \bibinfo {author} {\bibfnamefont {O.}~\bibnamefont {Bludov}},
		\bibinfo {author} {\bibfnamefont {V.}~\bibnamefont {Pashchenko}}, \bibinfo
		{author} {\bibfnamefont {S.~L.}\ \bibnamefont {Gnatchenko}}, \bibinfo
		{author} {\bibfnamefont {P.}~\bibnamefont {Lemmens}},\ and\ \bibinfo {author}
		{\bibfnamefont {H.}~\bibnamefont {Berger}},\ }\bibfield  {title} {\bibinfo
		{title} {{Magnetic properties of the antiferromagnetic spin-$\frac{1}{2}$
				chain system $\ensuremath{\beta}$-TeVO${}_{4}$}},\ }\href
	{https://doi.org/10.1103/PhysRevB.84.104447} {\bibfield  {journal} {\bibinfo
			{journal} {Phys. Rev. B}\ }\textbf {\bibinfo {volume} {84}},\ \bibinfo
		{pages} {104447} (\bibinfo {year} {2011})}\BibitemShut {NoStop}%
	\bibitem [{\citenamefont {Deen}\ \emph {et~al.}(2015)\citenamefont {Deen},
		\citenamefont {Florea}, \citenamefont {Lhotel},\ and\ \citenamefont
		{Jacobsen}}]{Deen014419}%
	\BibitemOpen
	\bibfield  {author} {\bibinfo {author} {\bibfnamefont {P.~P.}\ \bibnamefont
			{Deen}}, \bibinfo {author} {\bibfnamefont {O.}~\bibnamefont {Florea}},
		\bibinfo {author} {\bibfnamefont {E.}~\bibnamefont {Lhotel}},\ and\ \bibinfo
		{author} {\bibfnamefont {H.}~\bibnamefont {Jacobsen}},\ }\bibfield  {title}
	{\bibinfo {title} {{Updating the phase diagram of the archetypal frustrated
				magnet Gd$_{3}$Ga$_{5}$O$_{12}$}},\ }\href
	{https://doi.org/10.1103/PhysRevB.91.014419} {\bibfield  {journal} {\bibinfo
			{journal} {Phys. Rev. B}\ }\textbf {\bibinfo {volume} {91}},\ \bibinfo
		{pages} {014419} (\bibinfo {year} {2015})}\BibitemShut {NoStop}%
	\bibitem [{\citenamefont {Gopal}(2012)}]{Gopal2012}%
	\BibitemOpen
	\bibfield  {author} {\bibinfo {author} {\bibfnamefont {E.~S.~R.}\
			\bibnamefont {Gopal}},\ }\href@noop {} {\emph {\bibinfo {title} {{Specific
					Heats at Low Temperatures}}}}\ (\bibinfo  {publisher} {Springer},\ \bibinfo
	{address} {Boston, MA},\ \bibinfo {year} {2012})\BibitemShut {NoStop}%
	\bibitem [{\citenamefont {Magar}\ \emph {et~al.}(2022)\citenamefont {Magar},
		\citenamefont {Somesh}, \citenamefont {Singh}, \citenamefont {Abraham},
		\citenamefont {Senyk}, \citenamefont {Alfonsov}, \citenamefont {B\"uchner},
		\citenamefont {Kataev}, \citenamefont {Tsirlin},\ and\ \citenamefont
		{Nath}}]{Magar054076}%
	\BibitemOpen
	\bibfield  {author} {\bibinfo {author} {\bibfnamefont {A.}~\bibnamefont
			{Magar}}, \bibinfo {author} {\bibfnamefont {K.}~\bibnamefont {Somesh}},
		\bibinfo {author} {\bibfnamefont {V.}~\bibnamefont {Singh}}, \bibinfo
		{author} {\bibfnamefont {J.}~\bibnamefont {Abraham}}, \bibinfo {author}
		{\bibfnamefont {Y.}~\bibnamefont {Senyk}}, \bibinfo {author} {\bibfnamefont
			{A.}~\bibnamefont {Alfonsov}}, \bibinfo {author} {\bibfnamefont
			{B.}~\bibnamefont {B\"uchner}}, \bibinfo {author} {\bibfnamefont
			{V.}~\bibnamefont {Kataev}}, \bibinfo {author} {\bibfnamefont {A.~A.}\
			\bibnamefont {Tsirlin}},\ and\ \bibinfo {author} {\bibfnamefont
			{R.}~\bibnamefont {Nath}},\ }\bibfield  {title} {\bibinfo {title} {{Large
				Magnetocaloric Effect in the Kagome Ferromagnet
				Li$_9$Cr$_3$(P$_2$O$_{7}$)$_3$(PO$_4$)$_2$}},\ }\href
	{https://doi.org/10.1103/PhysRevApplied.18.054076} {\bibfield  {journal}
		{\bibinfo  {journal} {Phys. Rev. Appl.}\ }\textbf {\bibinfo {volume} {18}},\
		\bibinfo {pages} {054076} (\bibinfo {year} {2022})}\BibitemShut {NoStop}%
	\bibitem [{\citenamefont {Sebastian}\ \emph {et~al.}(2021)\citenamefont
		{Sebastian}, \citenamefont {Somesh}, \citenamefont {Nandi}, \citenamefont
		{Ahmed}, \citenamefont {Bag}, \citenamefont {Baenitz}, \citenamefont {Koo},
		\citenamefont {Sichelschmidt}, \citenamefont {Tsirlin}, \citenamefont
		{Furukawa},\ and\ \citenamefont {Nath}}]{Sebastian064413}%
	\BibitemOpen
	\bibfield  {author} {\bibinfo {author} {\bibfnamefont {S.~J.}\ \bibnamefont
			{Sebastian}}, \bibinfo {author} {\bibfnamefont {K.}~\bibnamefont {Somesh}},
		\bibinfo {author} {\bibfnamefont {M.}~\bibnamefont {Nandi}}, \bibinfo
		{author} {\bibfnamefont {N.}~\bibnamefont {Ahmed}}, \bibinfo {author}
		{\bibfnamefont {P.}~\bibnamefont {Bag}}, \bibinfo {author} {\bibfnamefont
			{M.}~\bibnamefont {Baenitz}}, \bibinfo {author} {\bibfnamefont
			{B.}~\bibnamefont {Koo}}, \bibinfo {author} {\bibfnamefont {J.}~\bibnamefont
			{Sichelschmidt}}, \bibinfo {author} {\bibfnamefont {A.~A.}\ \bibnamefont
			{Tsirlin}}, \bibinfo {author} {\bibfnamefont {Y.}~\bibnamefont {Furukawa}},\
		and\ \bibinfo {author} {\bibfnamefont {R.}~\bibnamefont {Nath}},\ }\bibfield
	{title} {\bibinfo {title} {{Quasi-one-dimensional magnetism in the
				spin-$\frac{1}{2}$ antiferromagnet
				${\mathrm{BaNa}}_{2}\mathrm{Cu}{({\mathrm{VO}}_{4})}_{2}$}},\ }\href
	{https://doi.org/10.1103/PhysRevB.103.064413} {\bibfield  {journal} {\bibinfo
			{journal} {Phys. Rev. B}\ }\textbf {\bibinfo {volume} {103}},\ \bibinfo
		{pages} {064413} (\bibinfo {year} {2021})}\BibitemShut {NoStop}%
	\bibitem [{\citenamefont {Golosovskii}\ \emph {et~al.}(1976)\citenamefont
		{Golosovskii}, \citenamefont {Plakhii}, \citenamefont {Smirnov},
		\citenamefont {Chernenkov}, \citenamefont {Kovalev},\ and\ \citenamefont
		{Bedrizova}}]{Golosovskii461}%
	\BibitemOpen
	\bibfield  {author} {\bibinfo {author} {\bibfnamefont {I.~V.}\ \bibnamefont
			{Golosovskii}}, \bibinfo {author} {\bibfnamefont {V.~P.}\ \bibnamefont
			{Plakhii}}, \bibinfo {author} {\bibfnamefont {O.~P.}\ \bibnamefont
			{Smirnov}}, \bibinfo {author} {\bibfnamefont {Y.~P.}\ \bibnamefont
			{Chernenkov}}, \bibinfo {author} {\bibfnamefont {A.~V.}\ \bibnamefont
			{Kovalev}},\ and\ \bibinfo {author} {\bibfnamefont {M.~N.}\ \bibnamefont
			{Bedrizova}},\ }\bibfield  {title} {\bibinfo {title} {{Magnetic ordering of
				Mn$^{2+}$ and Cr$^{3+}$ ions in the garnet
				$\mathrm{M}{\mathrm{n}}_{3}\mathrm{C}{\mathrm{r}}_{2}\mathrm{G}{\mathrm{e}}_{3}{\mathrm{O}}_{12}$}},\
	}\href {http://jetpletters.ru/ps/1815/article_27747.pdf} {\bibfield
		{journal} {\bibinfo  {journal} {JETP Lett.}\ }\textbf {\bibinfo {volume}
			{24}},\ \bibinfo {pages} {423} (\bibinfo {year} {1976})}\BibitemShut
	{NoStop}%
	\bibitem [{\citenamefont {Gukasov}\ \emph {et~al.}(1999)\citenamefont
		{Gukasov}, \citenamefont {Plakhty}, \citenamefont {Dorner}, \citenamefont
		{Kokovin}, \citenamefont {Syromyatnikov}, \citenamefont {Smirnov},\ and\
		\citenamefont {Chernenkov}}]{Gukasov2869}%
	\BibitemOpen
	\bibfield  {author} {\bibinfo {author} {\bibfnamefont {A.}~\bibnamefont
			{Gukasov}}, \bibinfo {author} {\bibfnamefont {V.}~\bibnamefont {Plakhty}},
		\bibinfo {author} {\bibfnamefont {B.}~\bibnamefont {Dorner}}, \bibinfo
		{author} {\bibfnamefont {S.~Y.}\ \bibnamefont {Kokovin}}, \bibinfo {author}
		{\bibfnamefont {V.}~\bibnamefont {Syromyatnikov}}, \bibinfo {author}
		{\bibfnamefont {O.}~\bibnamefont {Smirnov}},\ and\ \bibinfo {author}
		{\bibfnamefont {Y.~P.}\ \bibnamefont {Chernenkov}},\ }\bibfield  {title}
	{\bibinfo {title} {{Inelastic neutron scattering study of spin waves in the
				garnet with a triangular magnetic structure}},\ }\href
	{https://doi.org/10.1088/0953-8984/11/14/003} {\bibfield  {journal} {\bibinfo
			{journal} {J. Phys.: Condens. Matter}\ }\textbf {\bibinfo {volume} {11}},\
		\bibinfo {pages} {2869} (\bibinfo {year} {1999})}\BibitemShut {NoStop}%
	\bibitem [{\citenamefont {Islam}\ \emph {et~al.}(2020)\citenamefont {Islam},
		\citenamefont {Singh}, \citenamefont {Somesh}, \citenamefont {Mukharjee},
		\citenamefont {Jain}, \citenamefont {Yusuf},\ and\ \citenamefont
		{Nath}}]{Islam134433}%
	\BibitemOpen
	\bibfield  {author} {\bibinfo {author} {\bibfnamefont {S.~S.}\ \bibnamefont
			{Islam}}, \bibinfo {author} {\bibfnamefont {V.}~\bibnamefont {Singh}},
		\bibinfo {author} {\bibfnamefont {K.}~\bibnamefont {Somesh}}, \bibinfo
		{author} {\bibfnamefont {P.~K.}\ \bibnamefont {Mukharjee}}, \bibinfo {author}
		{\bibfnamefont {A.}~\bibnamefont {Jain}}, \bibinfo {author} {\bibfnamefont
			{S.~M.}\ \bibnamefont {Yusuf}},\ and\ \bibinfo {author} {\bibfnamefont
			{R.}~\bibnamefont {Nath}},\ }\bibfield  {title} {\bibinfo {title}
		{{Unconventional superparamagnetic behavior in the modified cubic spinel
				compound ${\mathrm{LiNi}}_{0.5}{\mathrm{Mn}}_{1.5}{\mathrm{O}}_{4}$}},\
	}\href {https://doi.org/10.1103/PhysRevB.102.134433} {\bibfield  {journal}
		{\bibinfo  {journal} {Phys. Rev. B}\ }\textbf {\bibinfo {volume} {102}},\
		\bibinfo {pages} {134433} (\bibinfo {year} {2020})}\BibitemShut {NoStop}%
	\bibitem [{\citenamefont {Sebastian}\ \emph {et~al.}(2022)\citenamefont
		{Sebastian}, \citenamefont {Islam}, \citenamefont {Jain}, \citenamefont
		{Yusuf}, \citenamefont {Uhlarz},\ and\ \citenamefont
		{Nath}}]{Sebastian104425}%
	\BibitemOpen
	\bibfield  {author} {\bibinfo {author} {\bibfnamefont {S.~J.}\ \bibnamefont
			{Sebastian}}, \bibinfo {author} {\bibfnamefont {S.~S.}\ \bibnamefont
			{Islam}}, \bibinfo {author} {\bibfnamefont {A.}~\bibnamefont {Jain}},
		\bibinfo {author} {\bibfnamefont {S.~M.}\ \bibnamefont {Yusuf}}, \bibinfo
		{author} {\bibfnamefont {M.}~\bibnamefont {Uhlarz}},\ and\ \bibinfo {author}
		{\bibfnamefont {R.}~\bibnamefont {Nath}},\ }\bibfield  {title} {\bibinfo
		{title} {{Collinear order in the spin-$\frac{5}{2}$ triangular-lattice
				antiferromagnet ${\mathrm{Na}}_{3}\mathrm{Fe}{({\mathrm{PO}}_{4})}_{2}$}},\
	}\href {https://doi.org/10.1103/PhysRevB.105.104425} {\bibfield  {journal}
		{\bibinfo  {journal} {Phys. Rev. B}\ }\textbf {\bibinfo {volume} {105}},\
		\bibinfo {pages} {104425} (\bibinfo {year} {2022})}\BibitemShut {NoStop}%
	\bibitem [{\citenamefont {Hopkinson}\ \emph {et~al.}(2007)\citenamefont
		{Hopkinson}, \citenamefont {Isakov}, \citenamefont {Kee},\ and\ \citenamefont
		{Kim}}]{Hopkinson037201}%
	\BibitemOpen
	\bibfield  {author} {\bibinfo {author} {\bibfnamefont {J.~M.}\ \bibnamefont
			{Hopkinson}}, \bibinfo {author} {\bibfnamefont {S.~V.}\ \bibnamefont
			{Isakov}}, \bibinfo {author} {\bibfnamefont {H.-Y.}\ \bibnamefont {Kee}},\
		and\ \bibinfo {author} {\bibfnamefont {Y.~B.}\ \bibnamefont {Kim}},\
	}\bibfield  {title} {\bibinfo {title} {Classical antiferromagnet on a
			hyperkagome lattice},\ }\href {https://doi.org/10.1103/PhysRevLett.99.037201}
	{\bibfield  {journal} {\bibinfo  {journal} {Phys. Rev. Lett.}\ }\textbf
		{\bibinfo {volume} {99}},\ \bibinfo {pages} {037201} (\bibinfo {year}
		{2007})}\BibitemShut {NoStop}%
	\bibitem [{\citenamefont {Kanamori}(1959)}]{Kanamori87}%
	\BibitemOpen
	\bibfield  {author} {\bibinfo {author} {\bibfnamefont {J.}~\bibnamefont
			{Kanamori}},\ }\bibfield  {title} {\bibinfo {title} {{Superexchange
				interaction and symmetry properties of electron orbitals}},\ }\href
	{https://doi.org/https://doi.org/10.1016/0022-3697(59)90061-7} {\bibfield
		{journal} {\bibinfo  {journal} {J. Phys. Chem. Solids}\ }\textbf {\bibinfo
			{volume} {10}},\ \bibinfo {pages} {87} (\bibinfo {year} {1959})}\BibitemShut
	{NoStop}%
	\bibitem [{\citenamefont {Florea}\ \emph {et~al.}(2017)\citenamefont {Florea},
		\citenamefont {Lhotel}, \citenamefont {Jacobsen}, \citenamefont {Knee},\ and\
		\citenamefont {Deen}}]{Florea220413}%
	\BibitemOpen
	\bibfield  {author} {\bibinfo {author} {\bibfnamefont {O.}~\bibnamefont
			{Florea}}, \bibinfo {author} {\bibfnamefont {E.}~\bibnamefont {Lhotel}},
		\bibinfo {author} {\bibfnamefont {H.}~\bibnamefont {Jacobsen}}, \bibinfo
		{author} {\bibfnamefont {C.~S.}\ \bibnamefont {Knee}},\ and\ \bibinfo
		{author} {\bibfnamefont {P.~P.}\ \bibnamefont {Deen}},\ }\bibfield  {title}
	{\bibinfo {title} {Absence of magnetic ordering and field-induced phase
			diagram in the gadolinium aluminum garnet},\ }\href
	{https://doi.org/10.1103/PhysRevB.96.220413} {\bibfield  {journal} {\bibinfo
			{journal} {Phys. Rev. B}\ }\textbf {\bibinfo {volume} {96}},\ \bibinfo
		{pages} {220413} (\bibinfo {year} {2017})}\BibitemShut {NoStop}%
	\bibitem [{\citenamefont {Pecharsky}\ and\ \citenamefont
		{Gschneidner~Jr}(1999)}]{Pecharsky44}%
	\BibitemOpen
	\bibfield  {author} {\bibinfo {author} {\bibfnamefont {V.~K.}\ \bibnamefont
			{Pecharsky}}\ and\ \bibinfo {author} {\bibfnamefont {K.~A.}\ \bibnamefont
			{Gschneidner~Jr}},\ }\bibfield  {title} {\bibinfo {title} {{Magnetocaloric
				effect and magnetic refrigeration}},\ }\href
	{https://doi.org/https://doi.org/10.1016/S0304-8853(99)00397-2} {\bibfield
		{journal} {\bibinfo  {journal} {J. Magn. Magn. Mater.}\ }\textbf {\bibinfo
			{volume} {200}},\ \bibinfo {pages} {44} (\bibinfo {year} {1999})}\BibitemShut
	{NoStop}%
	\bibitem [{\citenamefont {Singh}\ \emph {et~al.}(2022)\citenamefont {Singh},
		\citenamefont {Sarangi}, \citenamefont {Samal},\ and\ \citenamefont
		{Nath}}]{Singh111941}%
	\BibitemOpen
	\bibfield  {author} {\bibinfo {author} {\bibfnamefont {V.}~\bibnamefont
			{Singh}}, \bibinfo {author} {\bibfnamefont {S.~N.}\ \bibnamefont {Sarangi}},
		\bibinfo {author} {\bibfnamefont {D.}~\bibnamefont {Samal}},\ and\ \bibinfo
		{author} {\bibfnamefont {R.}~\bibnamefont {Nath}},\ }\bibfield  {title}
	{\bibinfo {title} {{Magnetic phase transition and magneto-elastic coupling in
				Fe$_{1+x}$Cr$_{2-x}$Se$_4$ ($x=0.0-0.50$)}},\ }\href
	{https://doi.org/https://doi.org/10.1016/j.materresbull.2022.111941}
	{\bibfield  {journal} {\bibinfo  {journal} {Mater. Res. Bull.}\ }\textbf
		{\bibinfo {volume} {155}},\ \bibinfo {pages} {111941} (\bibinfo {year}
		{2022})}\BibitemShut {NoStop}%
	\bibitem [{\citenamefont {Pecharsky}\ and\ \citenamefont
		{Gschneidner}(1999)}]{Pecharsky565}%
	\BibitemOpen
	\bibfield  {author} {\bibinfo {author} {\bibfnamefont {V.~K.}\ \bibnamefont
			{Pecharsky}}\ and\ \bibinfo {author} {\bibfnamefont {J.}~\bibnamefont
			{Gschneidner}, \bibfnamefont {K.~A.}},\ }\bibfield  {title} {\bibinfo {title}
		{{Magnetocaloric effect from indirect measurements: Magnetization and heat
				capacity}},\ }\href {https://doi.org/10.1063/1.370767} {\bibfield  {journal}
		{\bibinfo  {journal} {J. Appl. Phys.}\ }\textbf {\bibinfo {volume} {86}},\
		\bibinfo {pages} {565} (\bibinfo {year} {1999})}\BibitemShut {NoStop}%
	\bibitem [{\citenamefont {{V. Franco, J.S Blázquez, B. Ingale, and A.
				Conde}}(2012)}]{Franco305}%
	\BibitemOpen
	\bibfield  {author} {\bibinfo {author} {\bibnamefont {{V. Franco, J.S
					Blázquez, B. Ingale, and A. Conde}}},\ }\bibfield  {title} {\bibinfo {title}
		{The magnetocaloric effect and magnetic refrigeration near room temperature:
			Materials and models},\ }\href
	{https://doi.org/10.1146/annurev-matsci-062910-100356} {\bibfield  {journal}
		{\bibinfo  {journal} {Annu. Rev. Mater. Res.}\ }\textbf {\bibinfo {volume}
			{42}},\ \bibinfo {pages} {305} (\bibinfo {year} {2012})}\BibitemShut
	{NoStop}%
	\bibitem [{\citenamefont {Singh}\ \emph {et~al.}(2020)\citenamefont {Singh},
		\citenamefont {Bag}, \citenamefont {Rawat},\ and\ \citenamefont
		{Nath}}]{Singh6981}%
	\BibitemOpen
	\bibfield  {author} {\bibinfo {author} {\bibfnamefont {V.}~\bibnamefont
			{Singh}}, \bibinfo {author} {\bibfnamefont {P.}~\bibnamefont {Bag}}, \bibinfo
		{author} {\bibfnamefont {R.}~\bibnamefont {Rawat}},\ and\ \bibinfo {author}
		{\bibfnamefont {R.}~\bibnamefont {Nath}},\ }\bibfield  {title} {\bibinfo
		{title} {{Critical behavior and magnetocaloric effect across the magnetic
				transition in ${\mathrm{Mn}}_{1+x}{\mathrm{Fe}}_{4-x}{\mathrm{Si}}_{3}$}},\
	}\href {https://doi.org/10.1038/s41598-020-63223-0} {\bibfield  {journal}
		{\bibinfo  {journal} {Sci. Rep.}\ }\textbf {\bibinfo {volume} {10}},\
		\bibinfo {pages} {6981} (\bibinfo {year} {2020})}\BibitemShut {NoStop}%
	\bibitem [{\citenamefont {Law}\ \emph {et~al.}(2018)\citenamefont {Law},
		\citenamefont {Franco}, \citenamefont {Moreno-Ram{\'i}rez}, \citenamefont
		{Conde}, \citenamefont {Karpenkov}, \citenamefont {Radulov}, \citenamefont
		{Skokov},\ and\ \citenamefont {Gutfleisch}}]{Law2680}%
	\BibitemOpen
	\bibfield  {author} {\bibinfo {author} {\bibfnamefont {J.~Y.}\ \bibnamefont
			{Law}}, \bibinfo {author} {\bibfnamefont {V.}~\bibnamefont {Franco}},
		\bibinfo {author} {\bibfnamefont {L.~M.}\ \bibnamefont {Moreno-Ram{\'i}rez}},
		\bibinfo {author} {\bibfnamefont {A.}~\bibnamefont {Conde}}, \bibinfo
		{author} {\bibfnamefont {D.~Y.}\ \bibnamefont {Karpenkov}}, \bibinfo {author}
		{\bibfnamefont {I.}~\bibnamefont {Radulov}}, \bibinfo {author} {\bibfnamefont
			{K.~P.}\ \bibnamefont {Skokov}},\ and\ \bibinfo {author} {\bibfnamefont
			{O.}~\bibnamefont {Gutfleisch}},\ }\bibfield  {title} {\bibinfo {title} {A
			quantitative criterion for determining the order of magnetic phase
			transitions using the magnetocaloric effect},\ }\href
	{https://doi.org/10.1038/s41467-018-05111-w} {\bibfield  {journal} {\bibinfo
			{journal} {Nat. Comm.}\ }\textbf {\bibinfo {volume} {9}},\ \bibinfo {pages}
		{2680} (\bibinfo {year} {2018})}\BibitemShut {NoStop}%
	\bibitem [{\citenamefont {Midya}\ \emph {et~al.}(2010)\citenamefont {Midya},
		\citenamefont {Mandal}, \citenamefont {Das}, \citenamefont {Banerjee},
		\citenamefont {Chandra}, \citenamefont {Ganesan},\ and\ \citenamefont
		{Barman}}]{Midya142514}%
	\BibitemOpen
	\bibfield  {author} {\bibinfo {author} {\bibfnamefont {A.}~\bibnamefont
			{Midya}}, \bibinfo {author} {\bibfnamefont {P.}~\bibnamefont {Mandal}},
		\bibinfo {author} {\bibfnamefont {S.}~\bibnamefont {Das}}, \bibinfo {author}
		{\bibfnamefont {S.}~\bibnamefont {Banerjee}}, \bibinfo {author}
		{\bibfnamefont {L.~S.~S.}\ \bibnamefont {Chandra}}, \bibinfo {author}
		{\bibfnamefont {V.}~\bibnamefont {Ganesan}},\ and\ \bibinfo {author}
		{\bibfnamefont {S.~R.}\ \bibnamefont {Barman}},\ }\bibfield  {title}
	{\bibinfo {title} {{Magnetocaloric effect in HoMnO$_3$ crystal}},\ }\href
	{https://doi.org/10.1063/1.3386541} {\bibfield  {journal} {\bibinfo
			{journal} {Appl. Phys. Lett.}\ }\textbf {\bibinfo {volume} {96}},\ \bibinfo
		{pages} {142514} (\bibinfo {year} {2010})}\BibitemShut {NoStop}%
	\bibitem [{\citenamefont {Li}\ \emph {et~al.}(2012)\citenamefont {Li},
		\citenamefont {Nishimura}, \citenamefont {Hutchison}, \citenamefont {Qian},
		\citenamefont {Huo},\ and\ \citenamefont {NamiKi}}]{Lingwei152403}%
	\BibitemOpen
	\bibfield  {author} {\bibinfo {author} {\bibfnamefont {L.}~\bibnamefont
			{Li}}, \bibinfo {author} {\bibfnamefont {K.}~\bibnamefont {Nishimura}},
		\bibinfo {author} {\bibfnamefont {W.~D.}\ \bibnamefont {Hutchison}}, \bibinfo
		{author} {\bibfnamefont {Z.}~\bibnamefont {Qian}}, \bibinfo {author}
		{\bibfnamefont {D.}~\bibnamefont {Huo}},\ and\ \bibinfo {author}
		{\bibfnamefont {T.}~\bibnamefont {NamiKi}},\ }\bibfield  {title} {\bibinfo
		{title} {{Giant reversible magnetocaloric effect in ErMn$_2$Si$_2$ compound
				with a second order magnetic phase transition}},\ }\href
	{https://doi.org/10.1063/1.4704155} {\bibfield  {journal} {\bibinfo
			{journal} {Appl. Phys. Lett.}\ }\textbf {\bibinfo {volume} {100}},\ \bibinfo
		{pages} {152403} (\bibinfo {year} {2012})}\BibitemShut {NoStop}%
	\bibitem [{\citenamefont {Midya}\ \emph {et~al.}(2012)\citenamefont {Midya},
		\citenamefont {Khan}, \citenamefont {Bhoi},\ and\ \citenamefont
		{Mandal}}]{Midya132415}%
	\BibitemOpen
	\bibfield  {author} {\bibinfo {author} {\bibfnamefont {A.}~\bibnamefont
			{Midya}}, \bibinfo {author} {\bibfnamefont {N.}~\bibnamefont {Khan}},
		\bibinfo {author} {\bibfnamefont {D.}~\bibnamefont {Bhoi}},\ and\ \bibinfo
		{author} {\bibfnamefont {P.}~\bibnamefont {Mandal}},\ }\bibfield  {title}
	{\bibinfo {title} {{Giant magnetocaloric effect in magnetically frustrated
				EuHo$_2$O$_4$ and EuDy$_2$O$_4$ compounds}},\ }\href
	{https://doi.org/10.1063/1.4754849} {\bibfield  {journal} {\bibinfo
			{journal} {Appl. Phys. Lett.}\ }\textbf {\bibinfo {volume} {101}},\ \bibinfo
		{pages} {132415} (\bibinfo {year} {2012})}\BibitemShut {NoStop}%
	\bibitem [{\citenamefont {Midya}\ \emph {et~al.}(2016)\citenamefont {Midya},
		\citenamefont {Mandal}, \citenamefont {Rubi}, \citenamefont {Chen},
		\citenamefont {Wang}, \citenamefont {Mahendiran}, \citenamefont {Lorusso},\
		and\ \citenamefont {Evangelisti}}]{Midya094422}%
	\BibitemOpen
	\bibfield  {author} {\bibinfo {author} {\bibfnamefont {A.}~\bibnamefont
			{Midya}}, \bibinfo {author} {\bibfnamefont {P.}~\bibnamefont {Mandal}},
		\bibinfo {author} {\bibfnamefont {K.}~\bibnamefont {Rubi}}, \bibinfo {author}
		{\bibfnamefont {R.}~\bibnamefont {Chen}}, \bibinfo {author} {\bibfnamefont
			{J.-S.}\ \bibnamefont {Wang}}, \bibinfo {author} {\bibfnamefont
			{R.}~\bibnamefont {Mahendiran}}, \bibinfo {author} {\bibfnamefont
			{G.}~\bibnamefont {Lorusso}},\ and\ \bibinfo {author} {\bibfnamefont
			{M.}~\bibnamefont {Evangelisti}},\ }\bibfield  {title} {\bibinfo {title}
		{{Large adiabatic temperature and magnetic entropy changes in
				$\mathrm{EuTi}{\mathrm{O}}_{3}$}},\ }\href
	{https://doi.org/10.1103/PhysRevB.93.094422} {\bibfield  {journal} {\bibinfo
			{journal} {Phys. Rev. B}\ }\textbf {\bibinfo {volume} {93}},\ \bibinfo
		{pages} {094422} (\bibinfo {year} {2016})}\BibitemShut {NoStop}%
\end{thebibliography}

%

\end{document}